\documentclass[a4paper,fleqn,usenatbib]{mnras}

\usepackage{newtxtext,newtxmath}

\usepackage[T1]{fontenc}
\usepackage{ae,aecompl}

\usepackage{graphicx}	
\usepackage{amsmath}	
\usepackage{amssymb}	

\numberwithin{equation}{section}

\title{Model atmospheres of sub-stellar mass objects} 

\author[I. Hubeny]{
Ivan Hubeny$^{1,2}$\thanks{E-mail: hubeny@as.arizona.edu}
\\
$^{1}$Institute of Astronomy, University of Cambridge, Madingley Road, 
Cambridge CB3 0HA, UK\\
$^{2}$Permanent address: Steward Observatory, University of Arizona, 931 N, Cherry Ave, Tucson, AZ 85721, USA
}

\date{Accepted XXX. Received YYY; in original form ZZZ}

\pubyear{2016}

\begin{document}
\label{firstpage}
\pagerange{\pageref{firstpage}--\pageref{lastpage}}
\maketitle

\begin{abstract}
We present an outline of basic assumptions and governing
structural equations describing atmospheres of substellar mass
objects, in particular the extrasolar giant planets and brown dwarfs.
Although most of the presentation of the physical and numerical 
background is generic, details of the implementation pertain
mostly to the code {\sc CoolTlusty.}
We also present a review of numerical approaches and computer codes
devised to solve the structural equations, and make a critical evaluation 
of their efficiency and accuracy.
\end{abstract}




\section{Introduction}

There have been a number of theoretical studies dealing with constructing
model atmospheres of the sub-stellar mass objects (SMO), most notably
extrasolar giant planets  (EGP) and brown dwarfs. 
In the context of EGPs, the first self-consistent model atmospheres were
produced by Seager \& Sasselov (1998), followed by Goukenleuque et al.
(2000) and Barman et al. (2001). The first extended grid of EGP model atmospheres
was constructed by Sudarsky et al. (2003).
There have been many more theoretical studies afterward, but it is not our aim here 
to provide a historical review of the field. 

Most of the literature deals with
the properties of constructed models and with analyses of  observations.
However, the basic physical assumptions and the methodology of model
construction is usually covered only in short sections, usually referring
to other papers, or is sometimes lost in Appendices of otherwise application
minded papers.

Here, we intend to fill this gap, and provide a systematic overview of basic physical 
assumptions, structural equations, and numerical methods to solve them.
We also would like to clarify some previously confusing points, because researchers in the 
field of extrasolar giant planets come from both the planetary science and the stellar
atmosphere communities and use their respective traditional terminologies, sometimes
using the same term (e.g., the effective temperature, albedo, etc.) to mean a completely 
different concept.

Section 2 of this paper contains an outline of the basic assumptions and governing
structural equations describing an SMO atmosphere. Section 3 then reviews
the essential elements of the numerical methods used to solve the structural
equations without unnecessary approximations, and Section 4 deals with some 
important details of the numerical procedure.
Section 5 briefly discusses the topic of approximate, gray or pseudo-gray, models.
They are useful as initial models for a subsequent iterative scheme to solve the
structural equations exactly, as well as a pedagogical tool to understand the atmospheric
temperature structure. Finally, in Section 6, we discuss a comparison of the present scheme
to other modeling approaches. We also include several Appendices where some
technical details are described.

We stress that while Section 2 presents a general outline of the physical background
which is largely universal and is adopted by a number of approaches and computer codes,
the material presented in Sections. 3 and 4 pertains mostly to the code {\sc CoolTlusty} 
(Hubeny et al. 2003, Sudarsky et al . 2003) which was developed as a variant of the universal
stellar atmosphere code {\sc tlusty} (Hubeny 1988, Hubeny \& Lanz 1995), although
analogous or similar techniques are adopted in other codes, as is summarized in Section 6.

 
\section{Physical background}
\label{phys}

We will describe here a procedure to compute the so-called classical model atmospheres;
that is, plane-parallel, horizontally homogeneous atmospheres in hydrostatic 
and radiative (or radiative+convective) equilibrium. 

The basic physical framework employed to model the atmospheres of SMOs represents a
straightforward extension of the physical description used in the theory of stellar 
atmospheres. For a comprehensive
discussion and detailed description of the basic physics and numerics in the stellar context,
refer to Hubeny \& Mihalas (2014; in particular Chaps. 12--13, 16--18).
\subsection{Basic structural equations}
\label{phys_bas}
The basic structural equations are the hydrostatic equilibrium equation and 
the energy balance equation, Since radiation critically influences the energy 
balance, the radiative transfer equation has to be viewed as one of the basic
structural equations. These equations are supplemented by the equation of
state and the equations that define the absorption and emission coefficient
for radiation. We shall briefly discuss these equations below.
\subsubsection{Radiative transfer equation} 
For a time-independent, horizontally homogeneous atmosphere, 
possibly irradiated by an external source which is symmetric with
respect to the normal to the surface,
the radiative transfer equation is written as
\begin{equation}
\mu \frac{dI(\nu,\mu,z)}{dz} = -\chi(\nu,z) I(\nu,\mu,z) +\eta^{\rm tot}(\nu,\mu,z),
\end{equation}
where $I$ is the specific intensity of radiation defined such as
$I \cos\theta\, d\nu dt dS d\Omega$ is the energy of radiation having a frequency
in the range $(\nu, \nu+d\nu)$ going through an
elementary  surface $dS$ in an element of solid angle $d\Omega$ around direction
of propagation ${\bf n}$, with angle $\theta$ between the normal to the surface 
element $dS$, and ${\bf n}$, in time interval $dt$.  
In the plane-parallel geometry, the state parameters 
depend only on one geometrical coordinate, the depth in the 
atmosphere, and the specific intensity
depends only on the angle $\theta$; we use a customary notation $\mu\equiv\cos\theta$. 

Further, $\chi$ and $\eta^{\rm tot}$ are the total absorption and emission 
coefficients, respectively. They include both the thermal as well as the scattering 
processes -- see below.
Here we assume that there are no external forces and no macroscopic 
velocities, so the absorption coefficient does not depend on $\mu$.
The emission coefficient may still depend on direction; however, for an
isotropic scattering the emission coefficient is also
independent of $\mu$, $\eta^{\rm tot}(\nu,\mu,z) = \eta^{\rm tot}(\nu, z)$.

In the following, we denote a dependence on frequency through index $\nu$
and omit an indication of the dependence on depth. 
The total absorption coefficient, or {\em extinction coefficient}, is written as
\begin{equation}
\label{chi}
\chi_\nu = \kappa_\nu + s_\nu,
\end{equation}
where $\kappa_\nu$ is the coefficient of true absorption,  which correspond to 
a process during which an absorbed photon
is destroyed, while $s_\nu$ is is the scattering coefficient, 
corresponding to a
process which removes a photon from the beam, but re-emits it in a 
different direction\footnote{Generally, a scattering process may be non-coherent, in
which case an absorbed and a re-emitted photon may have different
frequencies, for instance during resonance scattering in spectral lines, or
in Compton scattering.  However, we will not consider these processes here and
assume a coherent scattering.}
We note that this coefficient is sometime denoted as $\sigma_\nu$, but we use the 
notation with $s$ to avoid a confusion with cross sections which we denote $\sigma$ --
see below.

The total emission coefficient is also given as a sum of thermal and scattering
contributions. The latter refers only to continuum scattering;
scattering. In the context of SMO model atmospheres, spectral lines are  
treated with complete frequency redistribution,  in which case the scattering term 
is in fact a part of the thermal emission coefficient. The continuum scattering part is 
usually treated separately from the thermal part,
and the ``thermal emission coefficient'' is usually called the ``emission
coefficient.'' Specifically, the total emission coefficient is written as
\begin{equation}
\label{etatot}
\eta_\nu^{\rm tot} = \eta_\nu + \eta_\nu^{\rm sc}.
\end{equation}
In the case of coherent isotropic scattering,
\begin{equation}
\label{etasc}
\eta_\nu^{\mathrm{sc}} = s_\nu J_\nu.
\end{equation}
For cold objects, brown dwarfs and exoplanets, one usually assumes
local thermodynamic equilibrium (LTE), in which case
\begin{equation}
\label{etalte}
\eta_\nu = \kappa_\nu B_\nu,
\end{equation}
where $B_\nu$ is the Planck function,
\begin{equation}
\label{planck}
B_\nu = \frac{2h\nu^3}{c^2} \frac{1}{\exp(h\nu/kT)-1},
\end{equation}
where $T$ is the temperature, and $h$, $k$, $c$ are the Planck constant, Boltzmann 
constant,  and the speed of light, respectively.

It is customary to introduce the optical depth,
\begin{equation}
\label{taudef}
d\tau_\nu = -\chi_\nu dz,
\end{equation}
and the source function
\begin{equation}
S_\nu = \frac{\eta_\nu^{\rm tot}}{\chi_\nu},
\end{equation}

In LTE, and for coherent isotropic scattering, the source function is given by
\begin{equation}
\label{slte}
S_\nu = \epsilon_\nu B_\nu + (1-\epsilon_\nu) J_\nu,
\end{equation}
where
\begin{equation}
\label{epsdef}
\epsilon_\nu = \frac{\kappa_\nu}{\chi_\nu}.
\end{equation}
The term $(1-\epsilon_\nu)$ is sometimes called a {\em single-scattering albedo}.  

The transfer equation now reads
\begin{equation}
\label{rtestd}
\mu \frac{dI_{\nu}(\mu)}{d\tau_\nu} = I_\nu(\mu) - S_\nu.
\end{equation}
Introducing the moments of the radiation intensity as
\begin{equation}
\label{moments}
[J_\nu, H_\nu, K_\nu] \equiv \frac{1}{2}\int_{-1}^1 I_\nu(\mu)[1,\mu,\mu^2]\, d\mu,
\end{equation}
the moment equations of the transfer equation read
\begin{equation}
\label{hmom}
\frac{dH_\nu}{d\tau_\nu} = J_\nu - S_\nu,
\end{equation}
and
\begin{equation}
\label{kmom}
\frac{dK_\nu}{d\tau_\nu} = H_\nu
\end{equation}
Combining Eqs. (\ref{hmom}) and (\ref{kmom}) one obtains a second-order
equation 
\begin{equation}
\label{rtevef}
\frac{d^2\! K_\nu}{d\tau_\nu^2} = J_\nu - S_\nu,
\end{equation}

When dealing with an iterative solution of the set of all
structural equations that specifically include the radiative transfer equation,
it is advantageous to introduce a form factor,
usually called the {\em (variable) Eddington factor}
\begin{equation}
\label{vef}
f_\nu = \frac{K_\nu}{J_\nu},
\end{equation}
and to write the second-order form as
\begin{equation}
\label{rte}
\frac{d^2 (f_\nu J_\nu)}{d\tau_\nu^2} = J_\nu - S_\nu.
\end{equation}
This equation contains only the mean intensity,
$J_\nu$, that depends on frequency and
depth, but not the specific intensity, $I_{\nu}(\mu)$, which is also 
a function of the polar angle $\theta$.
The Eddington factor is not known or given a priori, but is
computed in the formal solution of the transfer equation, and 
is held fixed during the subsequent iteration of the linearization procedure.
By the term ``formal solution'' we mean a solution of the transfer equation with 
{\em known} source function. It is done between two consecutive iterations of the 
iterative scheme, with current values of the state parameters.

We stress that introducing the Eddington factor does not represent an approximation.
Equation (\ref{rte}) is {\em exact} at the convergence limit. 
It should also be stressed that the Eddington factor technique offers some, but not
spectacular, advantages in solving the transfer equation for radiation 
intensities alone, because the computer
time for solving directly a linear, angle-dependent transfer equation, Eq. (\ref{rtestd}),
or solving a second-order equation (\ref{rte}) iteratively, is not very much different
unless one deals with a large number of directions.
However, its main strength lies in providing an efficient way of
solving simultaneously the radiative transfer equation together with other structural 
equations to determine 
the radiation intensity and other state parameters (temperature, density, etc.)
self-consistently.

The upper boundary condition is written as
\begin{equation}
\label{rte_ubc}
\left[\frac{d(f_\nu J_\nu)}{d\tau_\nu}\right]_0 = g_\nu J_\nu(0)
- H_\nu^{\rm{ext}},
\end{equation}
where $g_\nu$ is the surface Eddington factor defined by
\begin{equation}
\label{vefg}
g_\nu\equiv \frac{1}{2}\int_{0}^1 I_\nu(\mu,0)\mu\,d\mu \Big/ J_\nu(0),
\end{equation}
and
\begin{equation}
\label{hext}
H_\nu^{\rm{ext}}\equiv\frac{1}{2}\int_{0}^1 I_\nu^{\rm{ext}}(-\mu)\mu\,d\mu,
\end{equation}
where $I_\nu^{\rm{ext}} (-\mu)$ is the external incoming intensity at the top of
the atmosphere. 
Two features are worth stressing. First, the right-hand side of Eq. (\ref{rte_ubc})
can be written as $H_{\rm out} - H_{\rm in}$, that is, as a difference of the
outgoing and incoming flux at the top of the atmosphere. Second, the integral in
Eq. (\ref{vefg}) is evaluated only over the outgoing directions, but the 
definition of the surface Eddington factor $g$ contains the mean intensity $J$ which 
is defined through an integral over all, outgoing and incoming, directions.

The lower boundary condition is written similarly,
\begin{equation}
\label{rte_lbc}
\left[\frac{d(f_\nu J_\nu)}{d\tau_\nu}\right]_{\tau_{\rm{max}}}
= H_\nu^+ - \frac{1}{2} J_\nu,
\end{equation}
where $H_\nu^+ = \frac{1}{2}\int_0^1 I_\nu(\mu,\tau_{\mathrm{max}})\mu\, d\mu$.
The factor $1/2$ on the right-hand side of Eq. (\ref{rte_lbc}) could be replaced by
another Eddington factor analogous to $g_\nu$, but because the radiation field
at the lower boundary is essentially isotropic, this factor would be very close to $1/2$
anyway.
One typically assumes the diffusion approximation at
the lower boundary, in which case $I_\nu(\mu)=B_\nu+\mu(dB_\nu/d\tau_\nu)$, 
thus $H_\nu^+ = (1/2)B_\nu+(1/3)(dB_\nu/d\tau_\nu)$;
hence Eq. (\ref{rte_lbc}) is written as
\begin{equation}
\label{rte_lbc2}
\left[\frac{d(f_\nu J_\nu)}{d\tau_\nu}\right]_{\tau_{\mathrm{max}}}
=\left[\frac{1}{2}(B_\nu - J_\nu)+
\frac{1}{3}\frac{d B_\nu}{d\tau_\nu}\right]_{\tau_{\mathrm{max}}}.
\end{equation}

To compare this treatment of the radiative transfer equation to the approaches
usually used in the Earth or for the solar system planetary atmospheres, several
points are worth stressing:
\begin{enumerate}
\item All frequencies are treated at the same footing. There is no artificial separation
of frequencies into the ``solar'' (optical) region, in which the dominant mechanism of
photon transport is scattering, and the ``infrared'' region, in which the dominant mechanism
of transport is absorption and thermal emission of photons. 

\item External irradiation is treated simply, but at the same time exactly, as an upper boundary 
condition for the radiative transfer equation. No additional contribution of an attenuated
irradiation intensity is artificially added to the source function.

\item The transfer equation does not contain any assumptions about a division of
an atmosphere into a series of vertically homogeneous slabs, with constant properties
within a slab, as is often done in planetary studies. The transfer equation is discretized,
as shown explicitly in Appendix A, and a manner of discretization in fact stipulates a
behavior of the source function between the discretized grid points, 
in which it is determined exactly. 
For instance, a second-order form
of the transfer equation, Eq. (\ref{rte}), automatically yields a second-order accurate
numerical scheme, i.e. the solution of the transfer equation is exact for a piecewise 
parabolic form of the source function between the grid points.
\end{enumerate}

\subsubsection{Hydrostatic equilibrium equation} 

Under the conditions met in SMO atmospheres,
the radiation pressure is negligible, and the hydrostatic equilibrium 
equation is given simply as
\begin{equation}
\label{he}
\frac{dP}{dz} = -\rho g,  \quad {\rm or}\quad \frac{dP}{dm} = g,
\end{equation}
where $P$ is the gas  pressure, and $m$ the column mass,
\begin{equation}
\label{dm}
dm = -\rho\,dz,
\end{equation}
which is typically used (at least in stellar applications) as the basic depth coordinate.
Equation (\ref{he}) has a simple solution $P=mg$, so one can use either
$P$ or $m$ as a depth coordinate.
\subsubsection{Radiative equilibrium equation} 
In the convectively stable layers, the condition of energy balance is represented 
by the radiative equilibrium equation,
\begin{equation}
\label{re1}
\int_0^\infty\!\!\left(\chi_\nu J_\nu - \eta^{\rm tot}_\nu \right)\,d\nu =0,
\end{equation}
which states that no energy is being generated in, 
or removed from, an elementary volume in the atmosphere. In other words, 
the total radiation energy emitted in a given volume is exactly balanced
to the total energy absorbed. This form of the radiative equilibrium equation
is called the {\em integral form}. 

In view of Eqs. (\ref{chi}) - (\ref{etalte}), the term representing the net radiative
energy generation can be written as
\begin{equation}
\label{scatrem}
\int_0^\infty\!\! (\chi_\nu J_\nu - \eta_\nu^{\rm tot})\, d\nu =
\int_0^\infty\!\! (\kappa_\nu J_\nu - \eta_\nu)\, d\nu 
\end{equation}
because the scattering terms exactly cancel. 
Physically, Eq.~(\ref{scatrem}) states
that the coherent scattering, which represents a process of an absorption plus subsequent 
re-emission of a photon without a change of its energy, does not contribute to the 
energy balance.

As follows from Eq. (\ref{etalte}), in LTE one has
\begin{equation}
\label{relte}
\int_0^\infty\!\! (\kappa_\nu J_\nu - \eta_\nu)\, d\nu =
\int_0^\infty\!\! \kappa_\nu (J_\nu - B_\nu) \,d\nu =0,
\end{equation}
but we will use a general term in the following text.

Using Eq. (\ref{hmom}), the radiative equilibrium equation can also be
written as
\begin{equation}
\label{re2}
\int_0^\infty \frac{dH_\nu}{dz} d\nu = 0, 
\end{equation}
or, equivalently,
\begin{equation}
\label{re2a}
H\equiv \int_0^\infty H_\nu d\nu = {\rm const} \equiv \frac{\sigma_{\!R}}{4\pi} T_{\rm eff}^4,
\end{equation}
where $\sigma_{\!R}$ is the Stefan-Boltzmann constant,
and $T_{\mathrm{eff}}$ the effective temperature, which is a measure of
the total energy flux coming from the interior. It is one of the basic parameters
of the problem.

We stress that we use the term ``effective temperature'' as it is used in 
the stellar context. In the planetary studies, this term is traditionally used to describe
an equilibrium temperature of the upper layers of an irradiated atmosphere.
So, this term has in a sense an opposite meaning in these two fields:
in the stellar atmosphere terminology it describes the energy flux coming from
the interior, and, in view of Eq. (\ref{re2a}), the {\em net} flux flux passing through the
atmosphere,
while in the planetary terminology it reflects the energy flux coming from the
outside. More accurately, in the planetary terminology it describes the {\em outgoing} flux
which, in most cases, almost balances the flux coming from the outside
and which can be substantially larger than the net flux.

Equation (\ref{re2}) can be rewritten, using Eqs. (\ref{kmom})
and (\ref{vef}), as
\begin{equation}
\int_0^\infty \frac{d(f_\nu J_\nu)}{ d\tau_\nu}\,d\nu = 
\frac{\sigma_{\!R} }{ 4\pi}\,T_{\mathrm{eff}}^4,
\end{equation}
which is called a {\em differential form} of the radiative equilibrium equation.
Experience with computing model stellar atmospheres 
(e.g. Hubeny \& Lanz 1995) revealed that it is numerically advantageous to
consider a linear combination of both forms of the radiative equilibrium
equation, namely
\begin{equation}
\label{re}
\alpha\bigg[\int_0^\infty\!\!\left(\kappa_\nu J_\nu - \eta_\nu\right)d\nu\bigg] +
\beta \bigg[\int_0^\infty \frac{d(f_\nu J_\nu)}{ d\tau_\nu}\,d\nu -
\frac{\sigma_{\!R} }{ 4\pi}\,T_{\mathrm{eff}}^4\bigg] = 0,
\end{equation}
where $\alpha$ and $\beta$ are empirical coefficients that satisfy
$\beta\rightarrow 0$ in upper layers, and $\beta\rightarrow 1$ in deep
layers, while $\alpha\rightarrow 1$ in upper layers, and may be essentially
arbitrary elsewhere.

The reason for this treatment is the following: The condition of a constant 
total flux, $dH/dm=0$, or equivalently, 
$\int[d(f_\nu J_\nu)/d\tau_\nu]\,d\nu= (\sigma_{\!R}/4\pi) T_{\rm eff}^4$,
(the differential form), is accurate and numerically stable at deeper layers,
where the mean intensity and the flux change appreciably from depth to depth. 
Consequently, the derivatives with respect to optical depth are well constrained. 
In fact, it must be applied at the lower boundary in order to impose the condition 
for the total flux given through the effective temperature.

At low optical depths, the 
flux is essentially constant and moreover fixed by the conditions deeper in the
atmosphere (around monochromatic optical depths of the order of unity), so that
an evaluation of the derivatives is unstable, and often dominated by errors in the
current values of $\kappa_\nu$ and $J_\nu$. Moreover, the local temperature 
is constrained by this condition only indirectly.

The integral form, which is mathematically equivalent, schematically
written as $\int\kappa_\nu J_\nu d\nu=\int\kappa_\nu B_\nu d\nu$, is stable at all depths,
including low optical depths, and is directly linked to the local temperature through
the Planck function. It is applicable everywhere in the atmosphere..

\subsubsection{Radiative/convective equilibrium equation} 
An atmosphere is locally unstable against convection if the Schwarzschild 
criterion is satisfied,
\begin{equation}
\label{schw}
\nabla_{\rm rad} > \nabla_{\rm ad},
\end{equation}
where $\nabla_{\rm rad}= (d \ln T/d\ln P)_{\rm rad}$ is the logarithmic 
temperature gradient in radiative equilibrium, and $\nabla_{\rm ad}$ is the 
adiabatic gradient. The latter is viewed as a function of temperature and
pressure, $\nabla_{\rm ad}= \nabla_{\rm ad}(T,P)$. 
The density $\rho$ is considered to be a function of $T$ and
$P$ through the equation of state.

If convection is present, equation (\ref{re}) is modified to read
\begin{eqnarray}
\label{re_conv}
\alpha\bigg[\int_0^\infty\!\!\left(\kappa_\nu J_\nu - \eta_\nu\right)d\nu
+ \frac{\rho}{4\pi}\frac{dF_{\rm conv}}{dm}  \bigg] \nonumber \\ +
\beta \bigg[\int_0^\infty\! \frac{d(f_\nu J_\nu)}{ d\tau_\nu}\,d\nu -
\frac{\sigma_{\!R} }{ 4\pi}\,T_{\rm eff}^4
+ \frac{F_{\rm conv}}{ 4\pi}\bigg] = 0
\end{eqnarray}
where $F_{\rm conv}$ is the convective flux. Using the mixing-length
approximation, it is given by
[e.g.,  Hubeny \& Mihalas (2014; \S\,16.5]
\begin{equation}
\label{conv}
F_{\rm conv} = (gQH_{P}/32)^{1/2}(\rho c_P T)(\nabla-\nabla_{\rm el})^{3/2} (\ell/H_P)^2,
\end{equation}
where 
$H_P \equiv -(d\ln P/dz)^{-1} = P/(\rho g)$
is the pressure scale height, $c_P$ is the specific heat at constant pressure, and
$Q \equiv -(d\ln\rho/d\ln T)_P$. 
Further, $\ell/H_P$ is the ratio of the
convective mixing length to the pressure scale height, taken as a free parameter
of the problem. $\nabla$ is the actual logarithmic temperature gradient, and
$\nabla_{\rm el}$ is the gradient in the convective elements. The latter is 
determined by considering the efficiency of the convective transport; see,
e.g.,  Hubeny \& Mihalas (2014; \S\,16.5),
\begin{equation}
\label{nablae}
\nabla-\nabla_{\rm el} = (\nabla-\nabla_{\rm ad}) + {\cal B}^2/2 - 
{\cal B}\sqrt{{\cal B}^2/2 - (\nabla-\nabla_{\rm ad})},
\end{equation}
where
\begin{equation}
\label{convb}
{\cal B}= \frac{12\sqrt 2\, \sigma_{\!R} T^3}{\rho c_p (gQH_P)^{1/2} (\ell/H_P) }
\, \frac{\tau_{\rm el} }{ 1+ \tau_{\rm el}^{2}/2},
\end{equation}
and where $\tau_{\rm el} = \chi_{\!R} \ell$ is the optical thickness of the
characteristic convective element with size $\ell$. 

The gradient in the convective elements is thus a function of temperature,
pressure, and the actual gradient,
$\nabla_{\rm el} = \nabla_{\rm el}(T,P,\nabla)$. The convective flux
can also be viewed as a function of $T$, $P$, and $\nabla$.
It should be noted that although in many cases 
$\nabla \approx \nabla_{\rm ad}$, we do not enforce this relation explicitly.

\subsubsection{Equation of state} 

In the present context, the equation of state gives a relation between
density and pressure.
The gas pressure is given, assuming an ideal gas, by
\begin{equation}
\label{eos0}
P= kTN  = kT \sum_j N_j,
\end{equation}
and the mass density as
\begin{equation}
\label{eos1}
\rho = \sum_j N_j m_j = m_H \sum_j N_j \frac{m_j}{m_H} = 
\frac{\bar\mu m_H}{kT} P ,
\end{equation}
where $N$ is the total particle number density,
and $k$ the Boltzmann constant.  The total particle number density is
given by the sum of the number densities of the individual atomic or molecular
species, $N_j$; we assume that the number density of free electrons
is negligible. $m_j$ is the mass of the species $j$, $m_H$ the mass of the
hydrogen atom, and $\bar\mu$ the mean molecular weight, given by
\begin{equation}
\bar\mu = \frac{\sum_j N_j (m_j/m_H)}{\sum_j N_j}.
\end{equation}
The individual number densities (concentrations) $N_j$  are obtained by solving 
the chemical equilibrium equations, or possibly taking into account some
departures from chemical equilibrium (see \S\,\ref{nce}).

However, in an essentially
solar-composition cold gas, a majority of particles are the hydrogen molecules
and neutral helium atoms, in which case the mean molecular weight is simply
$\bar\mu=(1+4Y)/(0.5+Y)\approx 2.33$, where $Y\approx 0.1$ is the solar helium abundance 
(by number, with respect to hydrogen). Taking into account a contribution of
heavier elements, in particular C, N, O, a more reasonable (yet still
approximate) value is $\mu\approx 2.38$.

\subsubsection{Absorption and emission coefficients} 
The absorption coefficient is given by
\begin{eqnarray}
\label{kappa}
\kappa_\nu&=&
\sum_i \sum_\ell\sum_{u>\ell} n_{\ell,i} \sigma^{\rm line}_{\ell u}(\nu)
+ \sum_i N_i \sigma^{\rm cont}_{i}(\nu) \nonumber \\
&+& \sum_j N_j \sigma^{\rm cond, abs}_{j}(\nu)
+ \kappa_\nu^{\mathrm{add}},
\end{eqnarray}
where the first term represents the contribution of spectral lines, summed
over all species $i$, lower levels $\ell$ and upper levels $u$. The
second term is the contribution of continuum processes of species $i$.
Unlike the case of stellar atmospheres, these processes are not very
important in the case of SMO atmospheres, with the exception of the
collisional-induced absorption of H${}_2$.
The third term represents an absorption of photons on condensed particles,
and the last term a possible additional or empirical 
opacity not included in the previous terms.
In all cases, $\sigma(\nu)$ represents the corresponding cross section,
$N$ the corresponding number density, and $n$ the individual level population.
The correction for stimulated emission, $1-\exp(-h\nu/kT)$ is assumed to be included
in the transition cross sections. 

It should be stressed that cross sections for spectral lines describe line broadening
effects and thus depend on temperatures and appropriate perturber number densities;
the most important being the hydrogen molecule, H${}_2$, and atomic helium, He. 
Absorption cross sections for condensates
depend on assumed distribution of cloud particle sizes.
There are several distributions considered in the literature, most commonly used ones
being a lognormal distribution  (Ackerman \& Marley 2001), or a distribution 
given by Deirmendjian (1964), 
used by Sudarsky et al (2000, 2003), and subsequently  in all applications using 
the {\sc CoolTlusty} modeling code, 
\begin{equation}
\label{deir}
n(a) \propto (a/a_0)^6 \exp[-6(a/a_0)],
\end{equation}
where $a_0$ is the modal particle size, usually taken as a free parameter. The adopted
cross section is then a function of $a_0$, and is given by
\begin{equation}
\label{sigmaa0}
\sigma(a_0,\nu) = \int_0^\infty\!\! n(a) \sigma(a,\nu)\, da\, \Bigg/ \int_0^\infty\!\! n(a) da,
\end{equation}
where  $\sigma(a,\nu)$ is the cross section for absorption on condensates
of a single size, $a$, typically given by the Mie theory. 

\medskip

The scattering coefficient is given by
\begin{equation}
\label{kappasc}
s_\nu =  \sum_{i} N_{i} \sigma^{\mathrm{Ray}}_i(\nu)
+ \sum_{j} N_j \sigma^{\mathrm{cond,sc}}_j(\nu),
\end{equation}
where  $\sigma^{\mathrm{Ray}}_i$ is the Rayleigh scattering cross section 
of species $i$,  and $\sigma^{\mathrm{cond,sc}}_j$ is the cross section for 
Mie scattering on condensate species $j$. The same averaging as that expressed
by Eq. (\ref{sigmaa0}) is applied here as well.
Notice that the scattering and the absorption cross sections  
$\sigma^{\rm cond, sc}_{j}(\nu)$ and $\sigma^{\rm cond, abs}_{j}(\nu)$
are generally different.

The absorption coefficient (\ref{kappa} and the scattering coefficient (\ref{kappasc})
express the so-called opacities {\em per length}. They are measured in units of cm${}^{-1}$
(since cross sections are in cm${}^2$ and number densities in  cm${}^{-3}$).
In actual applications, one often
works in terms of opacities {\em per mass}, in units of cm${}^2$g${}^{-1}$. They are given
by, for instance for the total opacity,
\begin{equation}
\chi_\nu^\prime \equiv \chi_\nu/\rho.
\end{equation}
Since the particle number densities are roughly proportional to the mass density,
the opacity per mass is much less sensitive to the density than the opacity per
length. This property is used to advantage when constructing opacity tables,
because interpolating in density is more accurate using the opacity per mass.


\subsection{Treatment of external irradiation}
\label{irrad}

Assuming that the distance, $D$, between the star and the planet is
much larger than the stellar radius, $r_\ast$, then all the rays from the star
to a given point at the planetary surface are essentially parallel. 
The total energy received per unit area at the planetary surface at the
substellar point is (e.g., Hubeny \& Mihalas 2014, Eq. 3.72)
\begin{equation}
E = 2\pi (r_\ast/D)^2 \int_0^1 I_\ast(\mu)\,  \mu\,  d\mu = 
4\pi (r_\ast/D)^2\,  H_\ast = (r_\ast/D)^2 F_\ast\, ,
\end{equation}
where $H_\ast$ is the first moment of the specific intensity at the 
stellar surface, $H_\ast = (1/2) \int_{-1}^1 I_\ast(\mu)\, \mu\, d\mu
=(1/2) \int_0^1 I_\ast(\mu)\, \mu\, d\mu$ (the second equality is valid
if there is no incoming radiation at the stellar surface).
The incoming (physical) flux at the planetary surface, intercepted by an
area perpendicular to the line of sight toward the star (i.e., at the
substellar point) is thus given by
\begin{equation}
F^{\rm ext}_0 \equiv 2\pi \int_0^1 I^{\rm ext}\, \mu\, d\mu = E = 
4\pi (r_\ast/D)^2 H_\ast\, ,
\end{equation}
Expressing the intercepted flux as the first moment of the
specific intensity, $H^{\rm ext}_0 = F^{\rm ext}_0/4\pi$, then
\begin{equation}
H^{\rm ext}_0 = (r_\ast/D)^2 H_\ast\, ,
\end{equation}

If one does not compute separate model atmospheres for individual annuli
corresponding to different positions of a star on the planetary sky
(i.e., at different distances from the substellar point), and instead uses
some sort of averaging over the planetary surface, then one has to introduce
an additional parameter, $f$, that accounts for the fact that the planet has a
non-flat surface. If we assume that the incoming irradiation energy is evenly
distributed over the irradiated hemisphere, then $f=1/2$;
if we assume that the incoming energy is redistributed over the whole surface,
then $f=1/4$. Such an averaged incoming flux is thus given by
\begin{equation}
H^{\rm ext} = f\, H^{\rm ext}_0 = f\, (r_\ast/D)^2 H_\ast\, .
\end{equation}

Finally, one needs to relate the incoming flux to the incoming specific
intensity because this is the quantity used for the upper
boundary condition for the transfer equation for specific intensity.
If we assume that the irradiation at the stellar surface is isotropic;
better speaking, we artificially isotropise a highly anisotropic
irradiation, $I^{\rm ext}(\mu) = I_0^{\rm ext}$, then
\begin{equation}
H^{\rm ext} = \frac{1}{2}\int_0^1 I^{\rm ext}(\mu)\mu\, d\mu =  \frac{1}{4} I_0^{\rm ext}, 
\end{equation}
and thus
\begin{equation}
I_0^{\rm ext} =  4 H_\ast (r_\ast/D)^2 f = \frac{F_\ast}{\pi} 
\left(\frac{r_\ast}{D}\right)^2\, f\, .
\end{equation}
This equation can be rewritten in a useful form, expressing $H_\ast = (\sigma_{\!R}/4\pi) 
T_\ast^4$.
where $T_\ast$ is the effective temperature of the irradiating star, as
\begin{equation}
\label{ibw}
I_0^{\rm ext} =  (\sigma_{\!R}/\pi) T_\ast^4 W = B(T_\ast) W,  
\end{equation}
where 
\begin{equation}
\label{dilw}
W \equiv (r_\ast/D)^2 f 
\end{equation}
is the so-called dilution factor. In the second equality in Eq. (\ref{ibw}), $B(T_\ast)$ is the total 
(frequency-integrated) Planck function.

\subsubsection{Day/night side interaction}

The above described formalism applies for any type of object that is irradiated
from an external source, such as a planet, a brown dwarf, or even a star in a
close binary system. Close-in planets that exhibit a tidally-locked
rotation present a special case. Their day and night sides exhibit a vastly
different atmospheric conditions, and therefore it is quite natural that 
an interaction of the day and the night side is important. A proper description
of this effect requires a hydrodynamic simulations (e.g., Komacek \& Showman 
2016, and references therein) and is thus
beyond the scope of simple atmospheric models considered here. However,
there are several approaches suggested in the literature that deal with this
effect in an approximate way, which will be described below.

This simplest way, considered e.g. in Sudarsky et al. (2003).
is based on characterizing the degree of the day/night side heat redistribution 
through an empirical parameter $f$, as described above.
Burrows et al. (2006) introduced an analogous parameter, $P_n$, 
as a fraction of incoming flux 
that is redistributed to the night side. The underlying assumption is
that the fraction $P_n$ of the incoming flux is
somehow removed before the incoming radiation reaches the
upper boundary of the atmosphere, and is deposited at
the lower boundary of the night-side atmosphere.

A more realistic approach was suggested by Burrows et al. (2008). 
The day side of the planet is irradiated by the true external radiation
coming from the star, but then a fraction $P_n$ is being removed at a certain
depth range, parameterized by limiting pressures $P_0$ and $P_1$.
The same amount of energy is deposited at the night side,
also in a certain depth range,
usually but not necessarily in the same pressure range. The rationale
for this approach is that meridional circulations, that may occur below the
surface, may actually carry a significant amount of energy to the night side.

Specifically, the total radiation flux (expressed as $H$) received by a unit surface 
of a planet at the angular distance $\mu_0$ from the substellar point is given by
\begin{equation}
H_{\rm tot}^{\rm ext}(\mu_0) = \left(\frac{r_\ast}{D}\right)^2 \mu_0 
\int_0^\infty\!\! H^\ast_\nu d\nu = \left(\frac{r_\ast}{D}\right)^2 \mu_0 \frac{\sigma_{\!~R}}{4\pi} 
T_\ast^4,
\end{equation}
so that the integrated flux over the surface of the dayside hemisphere is
\begin{equation}
\bar H_{\rm tot}^{\rm ext} \equiv \int_0^1\! H_{\rm tot}^{\rm ext}(\mu_0)\, d\mu_0 =
\frac{1}{2} \left(\frac{r_\ast}{D}\right)^2 \frac{\sigma_{\!R}}{4\pi} T_\ast^4.
\end{equation}
One defines a local gain/sink of energy, $D(m)$, such that
\begin{equation}
\int_0^\infty D(m) = H^{\rm irr},
\end{equation}
where
\begin{equation}
H^{\rm irr} \equiv P_n \bar{H}_{\rm tot}^{\rm ext}.
\end{equation}
One assumes that $D(m)$ is non-zero only between column masses $m_0$ and $m_1$
defined through limiting pressures $P_0$ and $P_1$. These are free, essentially ad-hoc 
parameters that aim to mimic a complex radiation-hydrodynamical process. Hydro simulations
may in principle provide a guidance to the choice of these parameters. Burrows et al. (2008)
adopted as an educated guess the values $P_0=0.05$, $P_1=0.5$ bars. 
$D(m)$ is negative (better speaking, 
non-positive) on the day side, and is non-negative on the night side.

One is free to choose an actual form of function $D(m)$; Burrows et al (2008) considered
two models, (i)  $D(m)$ being constant between $m_0$ and $m_1$, i.e.,
$D(m)= H^{\rm irr}/(m_1-m_0)$, or (ii) a model with $D(m)$ linearly decreasing between
$m_0$ and $m_1$, in such a way the $D(m)$ reaches 0 at $m=m_1$; then
$D(m) = 2 H^{\rm irr}(m_1-m)/(m_1-m_0)^2$.

The radiative equilibrium equation then becomes: in the integral form
\begin{equation}
\int_0^\infty\!\!\! \kappa_\nu (J_\nu - B_\nu) = -D(m),
\end{equation}
and in the differential form
\begin{equation}
\frac{dH}{dm} = -D(m), \quad {\rm or} \quad
H(m)=\frac{\sigma_{\!R}}{4\pi} T_{\rm eff}^4 + \int_m^{m_1}\!\!\! D(m^\prime)\, dm^\prime.
\end{equation}
These equations are easily modified for the convection zone, in the case where
the gain/sink energy region overlaps the convection zone.


\subsection{Treatment of clouds} 
\label{clouds}

Ideally, the cloud properties, namely its position, extent, and 
a distribution of condensed particle sizes,
should be determined self-consistently with local atmospheric 
conditions. However, this is a very difficult problem which is not yet fully
solved, even in the context of cloud formation in the Earth
atmosphere. In the context of SMO atmospheres, one has to resort to
various approximations and  parameterizations of the problem.

Ackerman and Marley (2001) reviewed an earlier work, and developed
a simple, yet physically motivated treatment of cloud formation.
They formulate an equation for the mole fractions of the gas and condensed
phases of a condensable species, $q_g$ and $q_c$, respectively. This approach
sets the cloud base at depth $z$ where the $q_g(z)=q_s(z)$, 
where $q_s(z)$ is the vapor mole fraction corresponding to the saturation vapor 
pressure at depth $z$.. In other words, the cloud base is set at the point where 
the actual $T$-$P$ profile intersects the condensation curve of the species. Below
this point, there are no condensates,
\begin{equation}
\label{ambase}
q_c(z) = 0, \quad {\rm if} \quad q_g(z) < q_s(z),
\end{equation}
and above this point, where $q_g(z) \geq q_s(z)$,
the mole fraction of the condensate is given by an equation that expresses
a balance between turbulent diffusion that mixes both the gas and condensed particles
and transport them upward, and sedimentation that transport condensate downward,
\begin{equation}
\label{amfrac}
-K \frac{\partial(q_g+q_c)}{\partial z} - v_{\rm sed} q_c = 0, \end{equation}
where $v_{\rm sed}$ is the mass-weighted droplet sedimentation velocity, and
$K$ is the vertical eddy diffusion coefficient. The latter can be expressed, assuming
a free convection, as a function of basic state parameters (Ackerman \& Marley 2001), 
namely the atmospheric scale height, convective mixing length, mean molecular
weight, temperature, and density.
Sedimentation velocity is expressed as
\begin{equation}
\label{amsed}
v_{\rm sed} =  f_{\rm rain} v_{\rm conv}
\end{equation}
where  $f_{\rm rain}$, the ratio of the sedimentation 
velocity to the convective scale velocity, is taken as a free parameter of the problem.
For $f_{\rm rain} \rightarrow 0$, sedimentation is essentially disregarded,
which leads to a cloud extending from the base all the way upward. For
$f_{\rm rain} \gg 1$, sedimentation is very efficient, and the cloud mass distribution
exhibits a sharp, essentially exponential, decline above the base.

Equations (\ref{amfrac}) and (\ref{amsed}) apply in the convection zone. In
the convectively stable regions, one introduces two more free parameters,
a minimum ``mixing length'', and a minimum value of the $K$ coefficient, to be able 
to use the same expressions as in the convection zone. 

For the distribution of cloud particle sizes, Ackerman \& Marley (2001) 
assume a lognormal distribution, in which the geometric mean radius and
the number concentration of particles is expressed through $q_c$ and $f_{\rm rain}$,
so that it contains only one free  
parameter, the geometric standard deviation of the distribution.

Although the Ackerman-Marley model is physical motivated, it still inevitably contains
several adjustable free parameters. 
Alternatively, one can devise an approach that treats the cloud mass distribution parametrically,
but can mimic a cloud composed of several condensed species. It can also offer some
additional flexibility in treating cloud shapes (Sudarsky et al 2000, 2003, Burrows et al. 2006). 

This treatment of the clouds is based on the following simple model, which is 
also adopted in the {\sc CoolTlusty} code. 

The opacity (per gram of atmospheric material) of the given 
condensate $j$ at pressure $P$ s given by
\begin{equation}
\label{kmie}
\kappa^\prime_j(\nu, P) = {\cal N}_j M_j\, (A/\mu)\, S_j\, \bar k_j(\nu, a_{0,j})\, f_j(P)\, ,
\end{equation}
where ${\cal N}_j$ is the number density (mixing ratio) of the species $j$, 
$M_j$ its molecular weight, $\mu$ the mean molecular weight of the atmospheric 
material, $A$ the Avogadro number. Factor ${\cal N}_j M_j (A/\mu)$ transforms the
opacity per gram of condensate to the opacity per gram of atmospheric material.
$S_j$ is the supersaturation ratio,  $\bar k_j(\nu, a_{0,j})$ is the opacity
per gram of species $j$ at frequency $\nu$ and for the modal particle size $a_{0,j}$. 
{\sc CoolTlusty}, uses a previously computed 
table of $\bar k_j$ for a number of values of $a_0$  and frequencies $\nu$.
An analogous expression is used for the scattering opacity.

In Eq. (\ref{kmie}), the supersaturation ratio and the modal particle size are taken 
as free parameters of the model. Intrinsic optical properties of cloud particles
(i.e., the absorption and scattering coefficients) are contained in appropriate tables.
All the physics of cloud absorption and scattering is thus set up independently of the
model atmosphere code.

Cloud shape function is parametrized in the following way (Burrows et al. 2006):
The cloud base is set at pressure $P_0$, given typically as an intersection of the
current $T$-$P$ profile and the corresponding condensation curve. It can however 
be set differently -- see below. One also introduces a plateau region between this 
and a higher pressure, $P_1 \geq P_0$, which is meant to mimic a contribution
of other condensate species for which the given one serves as a surrogate. 
For a single isolated cloud, $P_1 \rightarrow P_0$, and the flat part would shrink
to a zero extent. However, for multiple cloud condensates, or for a convective
regions with multiple $T$-$P$ intersection points, it is advantageous to introduce
a flat part that mimics these phenomena.
On both sides of the flat part, $f$ decreases as a power low whose exponents
are free parameters of the problem. The cloud shape function is thus given by
\begin{equation}
\label{cloudsh}
f(P) = 
\left\{ \begin{array}{ll}
(P/P_0)^{c_0}, & P\leq P_0, \\ [4pt]
1,  & P_0 \leq P \leq P_1, \\ [4pt]
(P/P_1)^{-c_1}, & P\geq P_1,
\end{array} \right. ,
\end{equation}
In this model, the supersaturation ratio $S$ and the modal particle size $a_o$
are taken as free parameters.  The cloud shape function contains three more
free parameters, $P_1$, $c_0$, and $c_1$.


\subsection{Departures from chemical equilibrium}
\label{nce}

There are two kinds of departures from chemical equilibrium that are taken
into account in a number of studies of SMO atmospheres:
\begin{enumerate}
\item Departures due to the rainout of a condensable species. Burrows \&
Sharp (1999) developed a simple and useful procedure to treat such
departures from chemical equilibrium. The concentrations of the species
that are influenced by a rainout depend only on temperature and pressure,
and therefore one may construct corresponding opacity tables independently
of an actual model atmosphere. In other words, such departures from strict
chemical equilibrium lead only to a modification of the opacity table, but not
to a necessity to change a computational algorithm of constructing model 
atmospheres, in contrast to the next case, described below.

\item The second type of departures occurs in the case when the chemical 
reaction time for certain important reactions is much larger than vertical transport
(mixing) timescale. The mechanism is sometimes referred to as ``quenching"
(for a recent review of the literature on the subject,  see Madhusudhan et al. 2016) 
It is usually considered for the carbon and
nitrogen chemistry. These are described schematically by the net reactions
\begin{equation}
\label{cchem}
{\rm CO} + 3{\rm H}_2 \longleftrightarrow {\rm CH}_4 + {\rm H}_2{\rm O},
\end{equation}
and
\begin{equation}
\label{nchem}
{\rm N}_2 + 3{\rm H}_2 \longleftrightarrow 2 {\rm NH}_4.
\end{equation}
Because of the strong C=C and N$\equiv$N bonds, the reactions
(\ref{cchem}) and (\ref{nchem}) proceed much faster form right to left than
from left to right. For instance, for carbon the reaction in which CO is converted
to CH${}_4$ is very slow, and therefore CO can be vertically transported 
by convective motions or eddy diffusion  to the upper and cooler atmospheric layers, 
in which it would be virtually absent in chemical equilibrium. The net
result is an overabundance of CO and N${}_2$ and an underabundance of
CH${}_4$ and NH${}_3$ in the upper layers of the atmosphere. The mechanism 
was first suggested by Prinn \& Barshay (1977) for the Jovian planets in the solar
system, and subsequently applied by Fegley \& Lodders (1996),
Griffith \& Yelle (1999) and Saumon et al. (2000) for the atmospheres of
brown dwarfs. Hubeny \& Burrows (2007) performed a systematic study of this
effect for the whole range of L and T dwarfs. We will use their notation and terminology
below.
\end{enumerate}

The mixing time is given by
\begin{equation}
t_{\rm mix} =
\left\{ \begin{array}{ll}
H^2 K_{zz}, & {\rm in\ the\ radiative\ zone},\\ [4pt]
3H_c/v_c, & {\rm in\ the\ convection\ zone},
\end{array}
\right.
\end{equation}
where $H$ is the pressure scale height, $K_{zz}$ is the coefficient of
eddy diffusion, $H_c$ the  convective mixing length (typically taken equal
to $H$), and $v_c$ is the convective velocity. While the mixing time in
the convective region is well defined, its value in the radiative region is
quite uncertain because of uncertainties in $K_{zz}$, which can attain
values between $10^2$ and $10^8$, as discussed, e.g.,  by  Saumon et al. 
(2006, 2007). 

The chemical time is also uncertain.
One can use the value of Prinn \& Barshay (1977) for carbon chemistry,
\begin{equation}
t_{\rm chem} \equiv t_{\rm CO}
= \frac{N({\rm CO})}{\kappa_{\rm CO} N({\rm H}_2) N({\rm H}_2{\rm CO})},
\end{equation}
with 
\begin{equation}
\kappa_{\rm CO} = 2.3\times 10^{-10} \exp(-36200/T),
\end{equation}
where $N({\rm A})$ is the number density of species A. Some other estimates of
the chemical time are available, see Hubeny \& Burrows (2007). For a more recent
treatment of non-equilibrium carbon chemistry, see, e.g., Visscher \& Moses (2011)
and Moses et al. (2011).

For nitrogen, the corresponding expressions are
\begin{equation}
t_{\rm chem} \equiv t_{{\rm N}_2}
= \frac{1}{\kappa_{{\rm N}_2} N({\rm H}_2) },
\end{equation}
with 
\begin{equation}
\kappa_{{\rm N}_2} = 8.54\times 10^{-8} \exp(-81515/T),
\end{equation}
For a more recent
treatment of non-equilibrium nitrogen chemistry, see, e.g., Moses et al. (2011).

The effects of departures of chemical equilibrium are treated in a simple way.
For the current $T$-$P$ profile, one finds an intersection point where the mixing time 
for the current $T$-$P$ profile equals the chemical reaction time. Above this point
(for lower pressures) the number densities of CO and CH${}_4$ are set to constant
values equal to those found at the intersection point. Analogous procedure is done 
for the nitrogen chemistry, fixing the N${}_2$ and NH${}_3$ number densities above
the intersection point. Since the amount of available oxygen atoms is changed by
this process (more are being sequestered by CO), the number density of water is also
held fixed above the intersection point.

\subsection{Empirical modifications of the basic equations}

\subsubsection{Modifications of radiative equilibrium} 

The radiative equilibrium equation (\ref{re}), or radiative/convective
equilibrium equation (\ref{re_conv}) can be modified by adding an empirical
energy loss/gain term, as was done foe instance by  Burrows et al. (2008).
One can introduce an empirical term $E(m)$, together with another parameter
$D(m)$ discussed in \S\,\ref{irrad}, so that the integral form of the
radiative equilibrium is written as
\begin{equation}
\int_0^\infty (\kappa_\nu J_\nu - \eta_\nu)\, d\nu = - D(m) - E(m),
\end{equation}
where $E(m)$ represents an energy gain $E>0$ or loss ($E<0$) per
unit volume. Quantity $D(m)$ is related to an empirical redistribution of
incoming radiation (as was done in Burrows et al 2008), while $E(m)$ refers 
to some unspecified empirical energy gain/sink. 
\subsubsection{Modifications of chemical equilibrium} 
There are several possible modifications of the chemical equilibrium: 
\begin{enumerate}
\item A simple modification for a rainout of the species after Sharp \& Burrows
(1997). 
\item Considering departures from chemical equilibrium due to quenching
for carbon and nitrogen chemistry, arising from long chemical timescales as
compared to dynamical timescales, as described above in \S\,\ref{nce}; 
\item Mixing ratios of the individual species can be set up completely
empirically, such as in Madhusudhan \& Seager (2009);
see also 
Line et al (2012), Madhusudhan et al. (2014); 
for a review refer to Madhusudhan et al. (2016).
In that case
the mixing ratios of selected species are treated as free parameters of
the problem.
\end{enumerate}

\subsubsection{Modifications of opacities} 

As indicated in Eq. (\ref{kappa}), one can include empirical opacity
sources. For instance, one may consider an artificial optical absorber
as in Burrows et al (2008) that represents an additional opacity source
in the optical region, placed at a certain depth range in the atmosphere.

\subsection{Synthetic (forward) versus analytic (retrieval) approach}

There are essentially two types of approaches to modeling atmospheres of
substellar-mass objects, and in particular the giant planets:
\begin{enumerate}
\item A {\em synthetic}, or {\em forward}, approach, in which one solves the basic 
structural equations to determine the structure of the atmosphere. computes 
a predicted spectrum, and compares the synthetic spectrum to observations.
When an agreement is consistently reached for the given set of basic input parameters 
of the model (effective temperature, surface gravity, chemical composition,
external irradiation, ), the analyzed object is declared to be described by the basic input
parameters equal to those of the model.  In this sense, one usually calls this procedure
a ``determination of the basic parameters.'' Another, perhaps even more important result
of such a study is that it verifies the validity of the basic physical picture of the studied
object.
This approach is exactly parallel to a usual approach in stelar physics where
one constructs a grid of model atmospheres together with synthetic spectra, and by
comparison to observations determines the basic input parameters of the model.

\item An {\em inverse}, or {\em retrieval} approach (also called or {\em analytic}, 
or {\em semi-empirical}  approach). Here one assumes a given structure
of the atmosphere. Typically, the temperature is assumed to be a prescribed
function of depth (pressure), and the chemical composition is either computed 
consistently with this $T$-$P$ profile, or is also set empirically. One then
computes emergent radiation for this atmosphere, and tries many such structures
until an agreement with observations is achieved. In the context of analysis of
exoplanets, this approach is usually called the retrieval' method (Madhusudhan \&
Seager 2009), also see Irwin et al. (2008), Line et al (2012, 2013), Madhusudhan et al. (2014), 
and  for a review refer to Madhusudhan et al. (2016).

\end{enumerate}

An advantage of the synthetic approach is that it computes a model based on
true physical and chemical description. But, the disadvantage is that the input 
physics and chemistry is often very uncertain or approximate. Thus the analytic
approaches have a potential to highlight missing parts of physics and chemistry.
As an example from a different field, semi-empirical models of the solar
atmosphere (e.g. Vernazza et al. 1973) showed that the radiative equilibrium
assumption cannot hold in the uppermost layers (the chromosphere), and some 
additional source of energy has to be invoked.
These models determined the temperature as a function of depth needed to explain the 
observed spectral features, and even estimated the amount of extra energy needed 
to produce such a temperature structure.

Here, we will mostly describe the synthetic approach, but will also  describe the
methods used to obtain the emergent radiation from the given structure,
which is at the heart of the analytic method.


\subsection{1-D versus multi-D models}
\label{multid}

The basic approximation inherent in the above described modeling approach is the 
assumption of a plane-parallel horizontally-homogeneous, i.e. a 1-dimensional
(1-D) atmosphere. In other words, the structural parameters are allowed to depend only 
on one coordinate -- the depth in the atmosphere.

There are several essential reasons why this approximation may be violated:
\begin{enumerate}
\item In the case of strong external irradiation, the atmospheric conditions 
depend on the angular distance of the given position in the atmosphere 
from the substellar point. 

\item If clouds of condensates are formed, they are most likely formed with an 
inhomogeneous distribution on the stellar/planetary surface.

\item For a close-in planet with a tidally-locked rotation period,
an interaction between the day and night sides will inevitable lead to meridional
circulations that may exhibit a rather complicated pattern (e.g., Komacek \& Showman 
2016).

\item The presence of convection leads to inhomogeneities,  but these typically occur
on small geometrical scales, so they are usually treated using horizontally-averaged 
(1-D) models.
\end{enumerate}

The first two issues may be dealt with approximately by using the concept of a
$1\scriptsize{\frac{1}{2}}$-D approach, in which one constructs a series of 
1-D models for individual patches of an atmosphere. 
\begin{enumerate}
\item In the case of strongly irradiated planets, one can construct models for rings (belts)
with an equal distance from the substellar point. In other words, all points on a given 
belt see the irradiating star at the same polar angle. This was actually done by
Barman et al. (2001). They found that the differences
between this approach and the original, fully 1-D one, are not big. Nevertheless,
for more accurate models these effects should be taken into account.

\item Similarly, one can deal with horizontal inhomogeneities due to clouds by constructing
1-D models with and without clouds. Introducing an empirical cloud-covering 
factor, $a$, one can approximate the predicted radiation from the object as 
\begin{equation}
F_\lambda = a F_\lambda^{\rm clouds} + (1-a) F_\lambda^{\rm no\ clouds}.
\end{equation}
One can also form a final spectrum by a linear combination of models with
various cloud extents, but in such a case the number of input empirical
parameters will become too large, with a questionable physical meaning.

\item To deal with inhomogeneities caused by meridional circulation and
other dynamical phenomena, the current approach is first to construct a
hydrodynamical model without radiation, or with a simplified treatment
of radiation transport (e.g., Showman \& Guillot 2002, Showman et al. 2009, 2010),
and using the atmospheric structure following from
such a model to compute ``snapshot'' spectra using detailed radiation transport,
possibly using methods described in the paper. This was done for instance
by Burrows et al. (2010)..
\end{enumerate}

One can in principle construct, using present computational facilities,  
more sophisticated 3-D radiation hydrodynamic model atmospheres of SMOs,
and in particular close-in exoplanets, but this field of study is still in its infancy.


\section{Numerical solution}

The set of structural equations (\ref{rte}), 
(\ref{rte_ubc}), (\ref{rte_lbc2}),  (\ref{re}) or (\ref{re_conv}),
and necessary auxiliary expressions, are discretized 
in depth and frequency, replacing derivatives by differences and integrals 
by quadrature sums. This yields a set of non-linear algebraic equations. 
Detailed forms of the discretized equations are summarized in Hubeny 
\& Mihalas (2014; \S\,18.1); see also Appendix A.

Upon discretization,
the physical state of an atmosphere is fully described by the set of vectors
$\psi_d$ for every depth point $d$, $(d=1, \ldots, N\!D)$, $N\!D$ being the total
number of discretized depth points. The full state vector ${\bf \psi}_d$ is given by
\vskip-4pt
\begin{equation}
\label{cl1}
{\bf \psi}_d = \{ J_1, \ldots, J_{N\!F}, T, [\rho] ,[\nabla]\},
\end{equation}
where $J_i$, $(i=1,\ldots, N\!F)$ is the mean intensity of radiation in the $i$-th frequency 
point; we have omitted the depth subscript $d$. $N\!F$ is the number of discretized
frequency points. The quantities in the square
brackets are optional, and are considered to be components of vector
$\psi$ only in specific cases. In most applications, $\rho$ and $\nabla$
are taken as function of $T$ and $P$. However, with the pressure $P$ being given
a priori as $P=mg$, they are viewed as functions of the temperature $T$ only.

\subsection{Linearization}
\label{cl}

Although the individual methods of solution may differ, the resulting set of
non-linear algebraic equations is solved by some kind of linearization.
Generally, a solution is obtained by an application of the Newton-Raphson method.
Suppose the required solution ${\bf\psi}_d$ can be written in terms of the current,
but imperfect, solution ${\bf\psi}_d^0$ as 
${\bf\psi}_d={\bf\psi}_d^0+{\bf\delta\psi}_d$.
The entire set of structural equations can be formally written as an operator
$P$ acting on the state vector ${\bf\psi}_d$ as 
\begin{equation}
\label{cl2}
P_d({\bf\psi}_d)=0.
\end{equation}  
To obtain the solution, we express 
$P_d({\bf\psi}_d^0+{\bf\delta\psi}_d)= 0$,
using a Taylor expansion of $P_d$,
\begin{equation}
\label{cl3}
P_d({\bf\psi}_d^0)
+\sum_j\frac{\partial P_d}{\partial\psi_{d,j}}{\bf\delta\psi}_{d,j} = 0,
\end{equation}
and solve for ${\bf\delta\psi}_d$. Because only the first--order (i.e., linear) term
of the expansion is taken into account, this approach is called a
\textit{linearization}. To obtain the corrections ${\bf \delta\psi}_d$, one has to
form a matrix of partial derivatives of all the equations with respect to all 
the unknowns at all depths---the \textit{Jacobi matrix}, or \textit{Jacobian}---and
to solve equation (\ref{cl3}).  The radiative equilibrium equation (in the differential 
form) couples two neighboring depth points $d\!-\!1$ and $d$, and the
radiative transfer equation couples depth point $d$ to two neighboring depths
$d\!-\!1$ and $d\!+\!1$; see equations (\ref{rte}) -- (\ref{rte_lbc}).
Consequently, the system of linearized equations can be written as
\begin{equation}
\label{cl4}
-{\bf A}_d {\bf \delta\psi}_{d-1} +{\bf B}_d {\bf \delta\psi}_{d} 
-{\bf C}_d {\bf \delta\psi}_{d+1} = {\bf L}_d,
\end{equation}
where ${\bf A}$, ${\bf B}$, and ${\bf C}$ are $N\!N \times N\!N$ matrices, 
with $N\!N$ being the dimension of vector $\psi_d$. The minus signs at the 
${\bf A}$ and ${\bf C}$ terms in Eq. (\ref{cl4}) are for convenience only.
The block of the first
$N\!F$ rows and $N\!F$ columns of any of matrices ${\bf A}$, ${\bf B}$, and ${\bf C}$ 
forms a diagonal sub-matrix (because there is no
coupling of the individual frequencies in the transfer equation), 
while the row and the column corresponding to $T$ are full (because the radiative 
or radiative/convective equilibrium equation
contains the mean intensity at all frequency points). 
${\bf L}$ is a residual error vector, given by 
\begin{equation}
\label{cl4a}
{\bf L}_d=-P_d({\bf \psi}_d^0). 
\end{equation}
At the convergence limit, $L\rightarrow 0$ and thus $\delta\psi_d\rightarrow 0$.

Equation (\ref{cl4}) forms a block-tridiagonal system, which 
is solved by a standard Gauss-Jordan elimination.
It consists of a forward elimination
\begin{equation}
\label{elim1}
{\bf D}_d = ({\bf B}_d - {\bf A}_d {\bf D}_{d-1})^{-1} {\bf C}_d,\quad d=2,\ldots,N\!D,
\end{equation}
starting with ${\bf D}_1 = {\bf B}_1^{-1} {\bf C}_1$; and 
\begin{equation}
\label{elim2}
{\bf Z}_d = ({\bf B}_d - {\bf A}_d {\bf D}_{d-1})^{-1} ({\bf L}_d + {\bf A}_d {\bf Z}_{d-1}) ,\quad d=2,\ldots,N\!D.
\end{equation}
with ${\bf Z}_1 = {\bf B}_1^{-1} {\bf L}_1$. The second part is
a back-substitution, 
\begin{equation}
\label{elim3}
{\bf \delta\psi}_d = {\bf D}_d  {\bf \delta\psi}_{d+1} +  {\bf Z}_{d} ,
\quad d=N\!D-1,\ldots,1,
\end{equation}
starting with ${\bf \delta\psi}_{\!N\!D} = {\bf Z}_{N\!D}$.

This procedure, known as {\it complete linearization}, was  developed in the seminal 
paper by Auer \& Mihalas (1969). However, one has to perform $N\!D$ inversions
of a $N\!N \times N\!N$ matrix per iteration -- see Eqs. (\ref{elim2}) and (\ref{elim3}).
Since the dimension of the state vector
$\psi$, that is, the total number of structural parameters $N\!N$ can be 
large; so unless the number of frequencies is very small (of the order
of few hundreds), a direct application of the original
complete linearization is too time consuming and therefore not practical. 

\subsection{Hybrid CL/ALI method}
\label{hybrid}

The method, developed by Hubeny \& Lanz (1995), combines the basic
advantages of the complete linearization (CL) and the accelerated lambda
iteration (ALI) method. 
We stress that this method employs just one
aspect of the general idea of the ALI schemes, expressed by Eq. (\ref{clali}) below.
More traditional applications of ALI provide an iterative solution of the radiative
transfer equation with a dominant scattering term in the
source function. One such application is outlined in \S\,\ref{ali}. 

The hybrid CL/ALI method is essentially the linearization method, with the only
difference from the traditional CL method being that
the mean intensity in some (most) frequency points is not treated as an
independent state parameter, but is instead expressed as
\begin{equation}
\label{clali}
J_{di} = \Lambda^\ast_{di} [\eta_{di}/\kappa_{di}] + \Delta J_{di},
\end{equation} 
where $d$ and $i$ represent indices of the discretized depth and frequency
points, respectively, $\Lambda^\ast$ is the so-called approximate Lambda
operator, and $\Delta J$ is a correction to the mean intensity. The approximate
operator is in most cases taken as a diagonal (local) operator, hence its action is just
an algebraic multiplication. It is evaluated in the formal solution of the transfer
equation, and is held fixed in the next iteration of the linearization procedure,
and so is the correction $\Delta J$. Since the absorption and emission
coefficients $\kappa$ and $\eta$ are known functions of temperature,
one may express the linearization correction to the mean intensity $J_{di}$ as
\begin{equation}
\label{clali2}
\delta J_{di} = \Lambda^\ast_{di} 
\frac{\partial (\eta_{di}/\kappa_{di} )}{\partial T_{di} }\delta T_{di},
\end{equation}

Equation (\ref{clali2}) shows that $J_{di}$ is effectively eliminated
from the set of unknowns, thus reducing
the size of vector $\psi$ to $N\!N = N\!F_{C\!L} + 1$, where
$N\!F_{C\!L}$ is the number of frequency points (called explicit frequencies)
for which the mean intensity is kept to be linearized. As was shown by
Hubeny \& Lanz (1995), $N\!F_{C\!L}$ can be very small, of the order
of $O(10^0)$ to a few times $10^1$. In the context of SMOs, 
this method was used for instance by
Sudarsky et al. (2003) to construct a grid of exoplanet model atmospheres.


\subsection{Rybicki scheme}
\label{ryb}

An alternative scheme, which can be used in conjunction with either the
original complete linearization, or with the hybrid CL/ALI scheme, is a generalization
of the method developed originally by Rybicki (1969) for solving a NLTE line transfer
problem. It starts with the same set of linearized structural equations, and consists 
of a reorganization of the state vector and the resulting Jacobi matrix in a
different form. Instead of forming a vector of all state parameters in a given
depth point, it considers a set of vectors of tmean intensity, each containing
the mean intensities in one frequency point for all depths,
\begin{equation}
\delta {\bf J}_i \equiv \{\delta J_{1i}, \delta J_{2i}, \ldots,\delta J_{N\!D,i} \},\quad
i=1,\ldots,N\!F,
\end{equation}
and analogously for the vector of temperatures
\begin{equation}
\delta {\bf T} \equiv \{\delta T_1, \delta T_2, \ldots,\delta T_{N\!D} \}.
\end{equation}
In a description of the method presented in Hubeny \& Mihalas (2014; \S\,17.3),
an analogous vector $\delta{\bf N}$ for the particle number density
was introduced, but this is not necessary here.

The linearized radiative transfer equation can be written as
\begin{equation}
\label{ryb1}
\sum_{d^\prime=d-1}^{d+1} U_{dd^\prime, i} \delta{\bf J}_{d^\prime i} +
\sum_{d^\prime=d-1}^{d+1} R_{dd^\prime, i} \delta{\bf T}_{d^\prime} = {E}_{di},
\end{equation}
for $i=1,\ldots,N\!F$.  In the matrix notation
\begin{equation}
\label{ryb1a}
{\bf U}_i \delta{\bf J}_i + {\bf R}_i \delta{\bf T} = {\bf E}_i ,
\end{equation}
where  ${\bf U}_i$ and ${\bf R}_i$ are $N\!D \times N\!D$
tridiagonal matrices that account for a coupling of the corrections to the
radiation field at frequency $\nu_i$ and the material properties that are taken as 
a function of $T$, at the three adjacent depth points $(d-1, d, d+1)$.

Analogously, the linearized radiative/convective equilibrium equation
is written as
\begin{equation}
\label{ryb2}
\sum_{i=1}^{N\!F}{\bf V}_i \delta{\bf J}_i + {\bf W} \delta{\bf T} = {\bf F} ,
\end{equation}
where ${\bf V}_i$ and ${\bf W}$ are generally bi-diagonal matrices
(in the differential form of the radiative/convective equilibrium equation; 
in the purely integral form they would be diagonal).

The overall structure here is reversed from the original complete linearization,
in the sense that the role of frequencies and depths is reversed.
The matrix elements are the same; they only appear in different places.
For instance,
\begin{equation}
U_{dd,i} \equiv (B_d)_{ii}, \quad U_{d,d-1,i} \equiv (A_d)_{ii}, 
\quad U_{d,d+1,i} \equiv (C_d)_{ii},
\end{equation}
\vspace{-2em}
\begin{equation}
R_{dd,i} \equiv (B_d)_{i,N\!F+1}, \quad R_{d,d-1,i} \equiv (A_d)_{i,N\!F+1},
\end{equation}
and so on.

The global system is a block-diagonal (since the frequency points are
not coupled), with an additional block (``row'') with the internal matrices
being tridiagonal. Corrections to the mean intensities are found from
Eq. (\ref{ryb1a}),
\begin{equation}
\label{ryb2a}
\delta{\bf J}_i = {\bf U}_i^{-1} {\bf E}_i - ({\bf U}_i^{-1} {\bf R}_i) \delta {\bf T}.
\end{equation}
Substituting Eq. (\ref{ryb2a}) into (\ref{ryb2}), one obtains for the correction 
of temperature
\begin{equation}
\label{ryb4}
\left( {\bf W} - \sum_{i=1}^{N\!F} {\bf V}_i {\bf U}_i^{-1} {\bf R}_i \right) 
\delta{\bf T} = 
\left( {\bf F} - \sum_{i=1}^{N\!F} {\bf V}_i {\bf U}_i^{-1} {\bf E}_i \right),
\end{equation}
which is solved for $\delta{\bf T}$., and then $\delta{\bf J}_i$ are obtained 
from Eq. (\ref{ryb2a}).

In this scheme, one has to invert $N\!F$ tridiagonal matrices ${\bf U}_i$,
which is very fast, plus one inversion of the
$N\!D \times N\!D$ grand matrix in Eq. (\ref{ryb4}), which is also fast.
Since the computer time scales linearly with the number of frequency
points, the method can be used even for models with a large number
of frequency points (several times $10^4$).
In the context of SMO's, this method was first used by 
Burrows et al. (2006) to construct a grid of L and T model atmospheres.
\begin{figure}
 \includegraphics[width=\columnwidth,height=5cm]{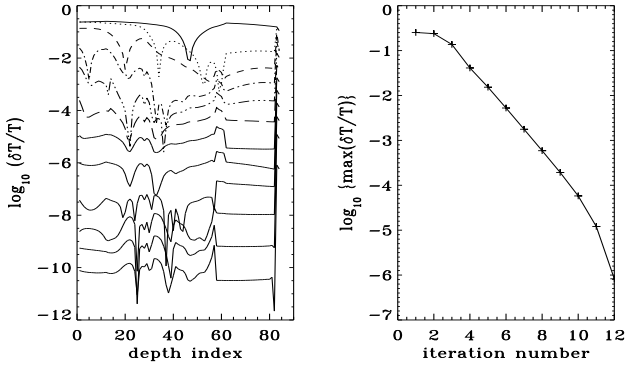}
 \caption{Illustration of the convergence properties of the Rybicki scheme.
 We display here a convergence log for a model atmosphere of a brown dwarf with 
 $T_{\rm teff}=1500$ K, $\log g = 5$; without clouds.
 Left panel: relative change of temperature,
 defined as $\delta T/T \equiv (T^{\rm new}- T^{\rm old})/T^{\rm old}$ as a function
 of depth, expressed as the depth index $d$. Here, $d=1$ corresponds to the
 uppermost point with the column mass $m_1=4.1\times 10^{-3}$ or pressure
 $P=4.1\times 10^{-4}$ bar, and $d=84$ corresponds to the 
 deepest point with $P=1.14\times 10^2$ bar, which correspond to the span of the
 Rosseland optical depths between $5.7\times 10^{-5}$ and $1.23\times 10^2$.
 The uppermost full line corresponds to the 1st iteration, dotted line to the 2nd,
 dashed line to the 3rd, and the subsequent lower lines correspond to the
 consecutive iteration steps. 
 The right panel displays the maximum relative change of temperature as a function
 of the iteration number. Both panels clearly demonstrate a very smooth and stable 
 convergence behavior of the Rybicki scheme.}
 \label{fig:f1}
\end{figure}

We illustrate the convergence properties of the Rybicki scheme on two
examples. First, we consider a brown dwarf model atmosphere computed with
{\sc CoolTtlusty}. Convergence pattern, displayed in Fig.\,\ref{fig:f1},
is similar to most of other SMO model atmosphere calculations.
Overall, the convergence properties are excellent.
The iteration process could have been safely stopped after the maximum
relative change of temperature decreased below $10^{-4}$; however we set the
convergence criterion here to be $10^{-5}$.

For the purposes of demonstration of numerical properties of the method, we 
chose a simplified numerical treatment with 5000 discretized frequency points
between $\nu = 6\times 10^{12}$ and $7\times 10^{14}$ s${}^{-1}$. Calculation
of the model took about 30 s on a MacBook Pro, OSX 10.9.5 with 2.2 GHz Intel i7 
processor, using an open-source {\tt gfortran} compiler.
We will show the properties of the actual model (temperature structure, conservation
of the total flux, numerical check of the radiative/convective equilibrium) later in
\S\,\ref{compar_plan}.

\begin{figure}
 \includegraphics[width=\columnwidth,height=5cm]{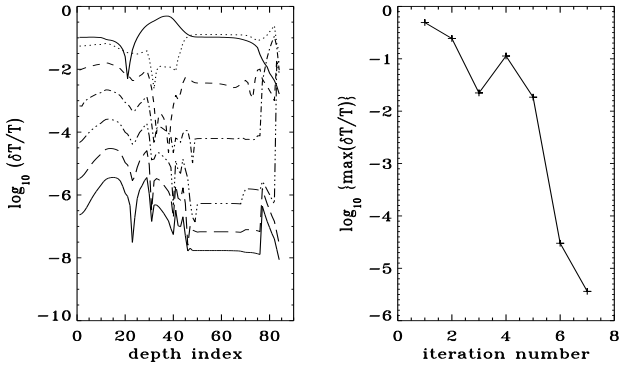}
 \caption{Convergence pattern for a model of a giant planet with $T_{\rm teff}=100$ K, 
 $\log g = 3$ irradiated by a solar-type star at a close distance of 0.06 A.U, computed 
 using the Rybicki scheme. The plot is analogous to Fig.\,\ref{fig:f1}. }
 \label{fig:fii2}
\end{figure}

\begin{figure}
 \includegraphics[width=\columnwidth,height=5cm]{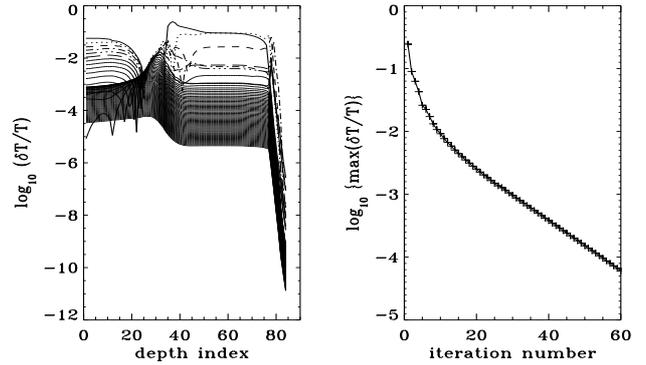}
 \caption{Convergence pattern for the same model as in Fig.\,\ref{fig:fii2},
 but computed with the hybrid CL/ALI method.}
 \label{fig:fii3}
\end{figure}

\begin{figure}
 \includegraphics[width=\columnwidth]{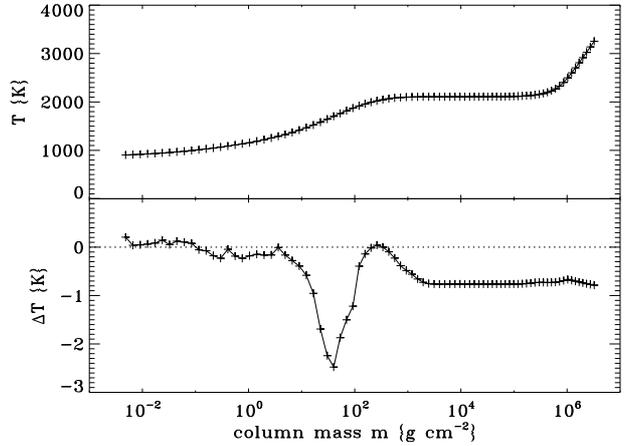}
 \caption{Temperature structure for the models displayed in Figs.\,\ref{fig:fii2} 
 and \ref{fig:fii3}.
 Upper panel: temperature as a function of the column density $m$. Lower panel:
 temperature difference between the two models.}
 \label{fig:fii4}
\end{figure}

Another example is a model atmospheres of a giant planet with $T_{\rm eff}=100$~K 
(in the stellar atmosphere terminology, i.e., with $T_{\rm eff}$ describing the total energy 
flux coming from the interior), $\log g = 3$,
irradiated by a solar-type star at a distance of 0.06 AU. The convergence
pattern is shown in Fig\,\ref{fig:fii2}. 
For comparison, we also show the convergence
pattern for the same model computed  using the hybrid CL/ALI method, where 
10 highest frequencies are treated using complete linearization, while the rest of 
frequencies are treated with  ALI -- see Fig.\,\ref{fig:fii3}. 
In order to be able to converge the model, one has to set the division parameters
$\alpha$ and $\beta$ in such a way that $\beta=1$ for $\tau_{\rm ross} \geq 0.5$,
and $\beta=0$ elsewhere, while $\alpha=1$ everywhere except the last 5 depth points
where it is set to 0.. 
Convergence is now much slower, although still stable. The corresponding
temperature structure is displayed in Fig.\,\ref{fig:fii4}. The upper panel shows the
temperature as a function of the column mass, while the lower panel shows the
temperature difference between the two models. Because the radiative/convective
equilibrium equation is solved differently in both cases, there are some differences,
albeit quite small and otherwise inconsequential.
%

\subsection{Overall procedure of the model construction}
\label{overall}

Construction of a model is composed of several basic steps,
which are described below.
\subsubsection{Initialization} 
Since  the overall scheme is an iterative one, an initial
estimate of a model is needed. It can be obtained in three possible ways:
\begin{enumerate}
\item Using a previously constructed model atmosphere for similar input
parameters. This way, one can compute a model with a  different chemical
composition, or with a slightly different irradiation flux than a model computed
earlier. If one does not change the input parameters significantly, the iterations
may proceed fast, and the overall computer time is shorter than when using other
methods for providing the initial model.
\item Using an LTE-gray model atmosphere. This is a typical method of obtaining
a starting model from scratch. The numerical procedure is described in Appendix C.
\item In some cases one can use an empirical temperature structure, using for
instance the parametric approach of Madhusudhan \& Seager (2009).
\end{enumerate}

\subsubsection{Global iteration loop} Each iteration consists of two main steps:
\begin{description}
\item [(A) {\it Formal solution.}] 
This step includes all calculation before entering any linearization step of the global scheme.
Take the current temperature, $T(m)$, and then:
\begin{enumerate}
\item Possibly smooth it if it exhibits a oscillatory behavior as a function of depth. 
\item Compute opacities (by interpolating in the opacity tables).  
\item Solve the radiative transfer equation for all frequency points -- see \S\,\ref{form}
and \S\,\ref{scat}. 
\item Recompute the temperature gradients (current and adiabatic), determine the position 
of the convection
zone, and possibly correct the temperature to satisfy the conservation of the total
(radiative + convective) flux -- \S\,\ref{convec}. 
\item With the new temperature, recalculate the mass density, and possibly return
to step (ii) and iterate several times. 
\end{enumerate}
This procedure results in a set of new values of structural parameters, $T$, $\rho$,
and $J_\nu$, which are as internally consistent as possible, and with which one
enters the next iteration of the global linearization scheme. This prudent procedure
increases the convergence speed and, in many cases, prevents convergence problems
or even a divergence of the global scheme.

\item [(B) {\it Linearization proper.}] This step includes evaluating the components of the
Jacobi matrix, and solving the global system, either for the corrections $\delta\psi$---when 
using the hybrid CL/ALI method (see \S\,\ref{hybrid}), or for ${\bf \delta T}$---when using 
the Rybicki scheme (see \S\,\ref{ryb}). As pointed out
above, the latter scheme is preferable. Using ${\bf \delta T}$, one evaluates the new
temperature structure $T(m)$, and returns to step {\rm (A)}.
\end{description}

We stress that the step (B), which may be called the ``temperature correction'',
should not be confused with  a procedure that is usually referred to by the same
name. The usual meaning of the term temperature correction is that it is a procedure
which employs the radiative/convective
equilibrium equation to update the local temperature to yield an improved
total energy flux, while keeping other parameters (radiation intensities, chemical 
composition, opacities) fixed. Here, step (B) indeed corrects the temperature, 
but {\em simultaneously} with other state parameters and the radiation intensities.
Consequently, the resulting convergence process is global and fast.


\section{Formal solution of the radiative transfer equation}
\label{forrte}

In the previous text, in particular in \S\,\ref{cl} -- \ref{ryb}, we have considered
a simultaneous solution of the transfer equation together with other structural
equations. To this end, we did not employ an angle-dependent transfer equation for
the specific intensity, but rather its combined moment equation for the mean 
intensity. Although such an equation is exact, it contains the Eddington factor
which is not known a priori, and which needs to be determined by a formal
solution of the (angle-dependent) transfer equation. 

By the term {\em formal solution} of the transfer equation we understand here
a determination of the specific intensity for a given absorption and (thermal)
emission coefficient.
There are several types of the formal solution; 
a detailed description of the most popular numerical schemes is presented
in Hubeny \& Mihalas (2014; \S\,12.4).


\subsection{Feautrier method}
\label{form}

If the source function is independent of $\mu$, as it is in the case of isotropic scattering,
or is an even function of $\mu$, then the most convenient method of the solution is the 
Feautrier (1964) method. It is based on introducing the symmetric and antisymmetric
averages of the specific intensity for $\mu\geq 0$,
\begin{eqnarray}
j_\nu(\mu) \equiv  [I_\nu(\mu) + I_\nu(-\mu)]/2,\\
h_\nu(\mu) \equiv [I_\nu(\mu) - I_\nu(-\mu)]/2.
\end{eqnarray}
Adding and subtracting the two forms of the transfer equation for
$\mu$ and $-\mu$, namely (suppressing the frequency index)
$\mu[dI(\mu)/d\tau] = I(\mu) -S$, and
$-\mu[dI(-\mu)/d\tau] = I(-\mu) -S$,
one obtains
\begin{equation}
\label{feah}
\mu\frac{d h_\nu(\mu)}{d\tau_\nu} = j_\nu(\mu) - S_\nu,\\
\end{equation}
and
\begin{equation}
\label{feaj}
\mu\frac{d j_\nu(\mu)}{d\tau_\nu} = h_\nu(\mu),
\end{equation}
and by differentiating Eq. (\ref{feaj}) once more and substituting into (\ref{feah}),
one obtains an exact equation for the symmetric average $j$, sometimes called the
Feautrier equation,
\begin{equation}
\label{feau}
\mu^2\, \frac{d^2 j_\nu(\mu)}{d\tau_\nu^2} = j_\nu(\mu) - S_\nu.
\end{equation}
It is interesting to point out that this scheme somewhat resembles the two-stream
approximation, often used in radiative transfer applications. However, unlike the
two-stream approaches, which are always approximate because they involve
some kind of averaging over one hemisphere, or representing one hemisphere 
by a single direction, the Feautrier equations  (\ref{feah}) - (\ref{feau}) are
{\em exact}.

Discretizing in the frequency and angle, and using Eq. (\ref{slte}) for
the source function, Eq. (\ref{feau}) becomes
\begin{equation}
\label{rtef}
\mu_i^2\,\frac{d^2 j_{ni}}{d\tau_n^2} = j_{ni} 
- (1-\epsilon_n) \sum_{i^\prime=1}^{N\!A} w_{i^\prime} j_{ni^\prime} - \epsilon_n B_n,
\end{equation}
where $N\!A$ is the number of angle points in one hemisphere, and
$w_i$ are the angular quadrature weights.

This equation is supplemented by the boundary conditions
\begin{equation}
\label{rtebc1}
\mu_i \left.\frac{dj_{ni}}{d\tau_n}\right|_0 = j_{ni}(0) - I_{ni}^{\rm ext},
\end{equation}
where $I_{ni}^{\rm ext}$ is the incoming specific intensity
$I(\nu_n, -\mu_i)$. The lower boundary condition reads
\begin{equation}
\label{rtebc2}
\mu_i \left.\frac{dj_{ni}}{d\tau_n}\right|_{\tau_{\rm max}}\!\! = I^+_{ni}(\tau_{\rm max})
 - j_{ni}(\tau_{\rm max}),
\end{equation}
where $I^+_{ni}(\tau_{\rm max})$ is the outward-defected specific intensity at
the deepest point, given by the diffusion approximation 
\begin{equation}
\label{rtebc3}
I^+_{ni}(\tau_{\rm max}) = B(\nu_n,\tau_{\rm max}) +
\mu_i\!\left.\frac{\partial B(\nu_n)}{\partial\tau_{\nu_n}}\right|_{\tau_{\rm max}},
\end{equation}

All the individual frequency points in Eqs. (\ref{rtef}) -- (\ref{rtebc3}) are independent,
so the transfer equation can by solved for one frequency at a time. We drop the  frequency
index $n$ and discretize in depth, described by index $d$.
Upon introducing a column vector ${\bf j}_d \equiv (j_{d,1}, J_{d,2},\ldots,j_{d,N\!A})$,
one writes Eqs. (\ref{rtef}) -- (\ref{rtebc3})  as a linear matrix equation
\begin{equation}
\label{rtefm} 
-{\bf A}_d {\bf j}_{d-1} +{\bf B}_d {\bf j}_{d} -{\bf C}_d {\bf j}_{d+1} = {\bf L}_d,
\end{equation}
where ${\bf A}_d$, ${\bf B}_d$,  and ${\bf C}_d$, are $N\!A \times N\!A$ matrices;
${\bf A}$ and ${\bf C}$ are diagonal, while ${\bf B}$ is full.
For illustration, we present here the matrix elements for the inner depth point
$d=2,\ldots,N\!D-1$; $i,j=1,\ldots, N\!A$,
\begin{eqnarray}
(A_d)_{ij}& =& \mu_i^2 /(\Delta\tau_{d-1/2,i} \Delta\tau_{d,i})\, \delta_{ij},\\
(C_d)_{ij}& =&\mu_i^2 /(\Delta\tau_{d+1/2,i} \Delta\tau_{d,i})\, \delta_{ij},\\
(B_d)_{ij}&=& (A_d)_{ij} + (C_d)_{ij} +\delta_{ij} - (1-\epsilon_d ) w_j
\end{eqnarray}
and
\begin{equation}
(L_d)_i = \epsilon_d B_d,
\end{equation}
where $\delta_{ij}$ is the Kronecker $\delta$-symbol, $\delta_{ij}=1$ for $i=j$
and $\delta_{ij}=0$ for $i\not=j$.
The expressions for the boundary conditions are analogous.

The system is solved by the standard Gauss-Jordan elimination, equivalent to
Egs. (\ref{elim1}) - (\ref{elim3}).
In terms of the Feautrier symmetric average $j$, the mean intensity and the Eddington factor are given by
\begin{equation}
\label{jdet}
J_d = \sum_{j=1}^{N\!A} w_j j_{dj}, \quad {\rm and}\quad
f_d = \sum_{j=1}^{N\!A} w_j \mu_j^2 j_{dj} \Big/ J_d.
\end{equation}

There are several  variants of the Feautrier scheme,
such as an improved second-order scheme by Rybicki \& Hummer (1991),
or a fourth-order Hermitian scheme by Auer (1976); for a detailed description
refer to Hubeny \& Mihalas (2014; \S\,12.3). 

All variants of the Feautrier method involve $N\!D$ inversions of 
$N\!A \times N\!A$ matrices. Since the typical value of
$N\!A$ is quite low (typically $N\!A=3$, which corresponds to 6
actual discretized angles), inverting such matrices does not present
any problem or any appreciable time consumption. The basic advantage
of the Feautrier scheme is that it treats scattering directly, without any
need to iterate.

It should be stressed that when using the Feautrier method for the formal
solution of the transfer equation between the subsequent iterations of the
global linearization scheme, one uses the above described procedure to
determine the Eddington factors. For consistency, one does not use the resulting
mean intensities directly, instead they are determined by solving Eq. (\ref{rte}), 
written as
\begin{equation}
\label{jrec}
\frac{d^2 (f_\nu J_\nu)}{d\tau_\nu^2} = \epsilon_\nu(J_\nu - B_\nu),
\end{equation}
because this is exactly the transfer equation as employed in the linearization
step. Otherwise the differences, albeit tiny, between $J_\nu$ determined from 
Eq. (\ref{jdet}) and from (\ref{jrec}) would prevent the overall iteration scheme 
to formally converge when using a very stringent convergence criterion,
because very near the converged solution the linearization would correct the mean 
intensities to satisfy Eq. (\ref{jrec}), while the formal solution through Eq. (\ref{jdet})
would change it back.


\subsection{Discontinuous Finite Element method}
\label{dfe}

If the source function depends on direction, or
if the number of angles is large (which may occur for some specific 
applications), or if an atmospheric structure exhibits very sharp variations 
with depth, it is advantageous to use the Discontinuous Finite Element (DFE) 
scheme by Castor et al. (1992). It solves the linear transfer equation (\ref{rtestd})
directly for the specific intensity, and therefore if scattering
is present, which is essentially always, the scattering part of the source
function has to be treated iteratively. To this end, a simple ALI-based procedure
is used. It is described, for a more complex case, below. Here we describe the
method assuming that the total source function is fully specified.

The method is essentially an application of the Galerkin method. The idea is to
divide a medium into a set of cells, and to represent the source function
within a cell by a simple polynomial, in this case by a linear segment.
The crucial point is that the segments are assumed to have
step discontinuities at grid points.  The specific intensity at grid point
$d$ is thus characterized by two values $I_d^+$ and $I_d^-$ appropriate
for cells $(\tau_d, \tau_{d+1})$ and $(\tau_{d-1}, \tau_d)$, respectively
(notice that we are dealing with an intensity in a given direction; 
the superscripts ``$+$'' and ``$-$'' thus
do not denote intensities in opposite directions as it is usually the case in
the radiative transfer theory).  The actual value of the specific intensity
$I(\tau_d)$ is given as an appropriate linear combination of $I_d^+$ and
$I_d^-$.
We skip all details here; suffice to say that after some algebra one
obtains simple recurrence relations for $I_d^+$ and $I_d^-$, 
for $d=1,\ldots,N\!D-1$,
\begin{eqnarray}
\label{dfem}
a_d I_{d+1}^- &=& 2 I_{d}^- + \Delta\tau_{d+1/2} S_{d} + b_d S_{d+1} , \\
\label{dfep}
a_d I_{d}^+ &=& 2(\Delta\tau_{d+1/2} + 1)\, I_{d}^- + b_d S_{d} - 
\Delta\tau_{d+1/2} S_{d+1},\ \ \ \ \ \ \ \ \ 
\end{eqnarray}
where
\begin{eqnarray}
\label{dfea}
a_d &=& \Delta\tau_{d+1/2}^2 + 2\Delta\tau_{d+1/2} + 2, \\
\label{dfeb}
b_d &=& \Delta\tau_{d+1/2} ( \Delta\tau_{d+1/2} +1),
\end{eqnarray}
and
\begin{equation}
\Delta\tau_{d+1/2} = (\tau_{d+1}-\tau_d)/|\mu|,
\end{equation}
which represents the optical depth differences along the line of photon propagation,
while $\tau$ measures the optical depth in the direction of the normal
to the surface.
The boundary condition is $I_1^-=I^{\rm ext}$, where $I^{\rm ext}$ is the specific
intensity of external irradiation (for inward-directed rays, $\mu<0$).

For outward-directed rays ($\mu>0$), one can either use the same expressions as above,
renumbering the depth points such as $N\!D \rightarrow 1, N\!D-1 \rightarrow 2,
\ldots,  1\rightarrow N\!D$; or to use the same numbering of depth points while setting
the recursion, for $d=N\!D-1,\ldots,1$, as
\begin{eqnarray}
a_d I_{d}^-\!\!\! &=& \!\!\!2 I_{d+1}^- + \Delta\tau_{d+1/2} S_{d+1} + b_d S_{d} , \\
\label{dfepm}
a_d I_{d+1}^+\!\!\! &=& \!\!\!2(\Delta\tau_{d+1/2}\! +\! 1)\, I_{d+1}^- + b_d S_{d+1} - 
\Delta\tau_{d+1/2} S_{d}, \ \ \ \ \ \ \ 
\end{eqnarray}
with $I_{d}^-=B_d+\mu (B_d-B_{d-1})/\Delta\tau_{d-1/2}$ for $d=N\!D$.

Finally, the resulting specific intensity at $\tau_d$ is given by a linear combinations
of the ``discontinuous" intensities $I_d^-$ and $I_d^+$ as
\begin{equation}
\label{dfei}
I_d = \frac{I_d^- \Delta\tau_{d+1/2}+ I_d^+ \Delta\tau_{d-1/2}}
{\Delta\tau_{d+1/2}+  \Delta\tau_{d-1/2}}.
\end{equation}
At the boundary points, $d=1$ and $d=N\!D$, we set $I_d = I_d^-$.
As was shown by Castor et al., it is exactly the linear combination of the
discontinuous intensities expressed by Eq. (\ref{dfei}) that makes the method 
second-order accurate. 
Since one does not need to evaluate any exponentials, the method is 
also very fast.  

We stress again that the above described scheme applies for a solution of the transfer
equation along a single angle of propagation. The source
function is assumed to be given. Therefore, when scattering is not negligible, one
has to iterate on the source function. This is done most efficiently using a very
powerful Accelerated Lambda Iteration (ALI) method, which will be outlined
in \S\,\ref{ali}.


\subsection{Anisotropic scattering on condensates}
\label{scat}

The scattering part of the emission coefficient is generally written as
\begin{equation}
\eta_\nu^{sc}({\bf n}) = s_\nu \oint (d\Omega^\prime/4\pi)\, I_\nu({\bf n}^\prime)\, 
g({\bf n}^\prime,{\bf n}),
\end{equation}
where $g({\bf n}^\prime,{\bf n})$ is the phase function for the scattering, ${\bf n}^\prime$
and ${\bf n}$ are the directions of the incoming and the scattered photon, respectively.
In the following text, the primed quantities refer to the incoming radiation and unprimed to
scattered radiation.

Introducing the usual polar ($\theta$) and the azimuthal ($\phi$) angles, with $\mu=\cos\theta$,
the source function with a general scattering term can be written as
\begin{equation}
\label{sfanis}
S(\nu,\mu,\phi) = \frac{1-\epsilon_\nu}{4\pi}\!\!\! \int_{-1}^1\!\!\!d\mu^\prime\!\!
\int_0^{2\pi}\!\!\!\! d\phi^\prime
I(\nu,\mu^\prime\!,\phi^\prime)\, g(\nu,\mu^\prime\!,\phi^\prime\!,\mu,\phi) + \epsilon_\nu B_\nu.
\end{equation}
The transfer equation to be solved is written as 
\begin{equation}
\label{rtean1}
\mu\frac{d I(\mu,\phi)}{d\tau} = I(\mu,\phi) - S(\mu,\phi).
\end{equation}
Here, and in the following expressions, we omit an explicit indication of the dependence
on frequency.
In general, Eq. (\ref{rtean1}) is not advantageous to be considered in the second-order form,
so the first-order form is solved, using the Discontinuous Finite Element method.\footnote{One
can also use the short characteristics method (e.g., Hubeny \& Mihalas 2014, \S\,12.4), but 
we will not consider this scheme here.} 

In the absence of external forces, the phase function depends only on the scattering
angle, that is the angle between the directions of the incoming and scattered photon,
which we denote as $\gamma$, where $\cos\gamma={\bf n}^\prime\cdot{\bf n}$.
In terms of the polar and azimuthal angles,
\begin{equation}
\cos\gamma= \sin\theta^\prime \sin\theta\, (\cos\phi^\prime \cos\phi + \sin\phi^\prime \sin\phi)
+\cos\theta^\prime \cos\theta.
\end{equation}

The simplest approximation is to treat both types of scattering that we deal with here, 
namely the Rayleigh and the Mie scattering, 
as being isotropic. In this case the phase function is simply 
\begin{equation}
g(\gamma) =1,
\end{equation}
and the source function is written in the usual form
\begin{equation}
S_\nu = (1-\epsilon_\nu) J_\nu + \epsilon_\nu B_\nu.
\end{equation}

For the Rayleigh scattering, one can either assume isotropic scattering, which is a crude
but acceptable approximation, or use an exact phase function which in this case is 
given by the dipole phase function,
\begin{equation}
g(\gamma) = \frac{3}{4}\,(1+\cos^2\gamma).
\end{equation}

For a scattering on cloud particles (condensates), there are three possible
approaches:
\begin{enumerate}
\item Assuming the isotropic phase function. This is a rough approximation, but is
acceptable for simple models, in particular when external irradiation is weak or
absent.
\item Employing the  Henyey-Greenstein phase function,
\begin{equation}
g(\gamma) =\frac{1- \bar g^2}{(1+\bar g^2-2\bar g\cos\gamma)^{3/2}},
\end{equation}
where $\bar g$ is the asymmetry parameter that is coming from the Mie theory.
\item Finally, the most accurate treatment is using an exact phase function that follows
from the Mie theory.
\end{enumerate}

In the two latter cases, one solves the transfer equation iteratively. One introduces a
form factor, analogous to the Eddington factor, as (see Sudarsky et al. 2005)
\begin{equation}
\label{adef}
a_{\mu\phi} = \frac{
\int_{-1}^1d\mu^\prime\int_0^{2\pi} d\phi^\prime I(\mu^\prime,\phi^\prime)\, 
g(\mu^\prime,\phi^\prime,\mu,\phi) }{4\pi J}.
\end{equation}
Notice that for isotropic scattering, $a_{\mu\phi}=1$. The iteration scheme proceeds as 
follows:
\begin{enumerate}
\item Initialize $a_{\mu\phi}$, usually as $a_{\mu\phi}=1$.
\item While holding $a_{\mu\phi}$ fixed, solve the transfer equation
with the source function given by
\begin{equation}
\label{sfan}
S_{\mu\phi} = (1-\epsilon) a_{\mu\phi} J + \epsilon B,
\end{equation}
for all angles $\mu$ and $\phi$, This can be done by the procedure described below.
\item After this is done, update $a_{\mu\phi}$, and repeat.
\end{enumerate}
In the absence of strong irradiation the radiation field is essentially independent of the
polar angle, so one can use a simpler procedure where the phase function is averaged
over azimuthal angles,
\begin{equation}
g(\mu^\prime,\mu) = \int_0^{2\pi} g(\mu^\prime,\mu,\phi^\prime, \phi_0) \, d\phi^\prime,
\end{equation}
where $\phi_0$ is an arbitrary value of the polar angle, typically chosen $\phi_0=0$.
The integration is performed numerically. The above equations are modified
correspondingly, essentially omitting the dependences on the polar angle.

The transfer equation is now
\begin{equation}
\label{rtean2}
\mu\frac{d I(\mu)}{d\tau} = I(\mu) - S(\mu),
\end{equation}
which can be put into the form involving the symmetric and antisymmetric averages,
analogous to the Feautrier scheme, namely
\begin{equation}
\label{rteanh}
\mu\frac{dh(\mu)}{d\tau} = j(\mu) -s\int_{-1}^1 g^+(\mu^\prime\!,\mu) j(\mu^\prime) d\mu^\prime,
\end{equation}
and 
\begin{equation}
\label{rteanj}
\mu\frac{dj(\mu)}{d\tau} = h(\mu) -s\int_{-1}^1 g^-(\mu^\prime\!,\mu) h(\mu^\prime) d\mu^\prime,
\end{equation}
where
\begin{equation}
g^{\pm}(\mu^\prime\!,\mu) = \frac{1}{2} [g(\mu^\prime\!,\mu) \pm g(\mu^\prime,-\mu)],
\end{equation}
because the following symmetry relations hold:
\begin{eqnarray} 
g(\mu^\prime\!,\mu) = g(-\mu^\prime\!,-\mu),\\
g(\mu^\prime\!,-\mu) = g(-\mu^\prime\!,\mu).
\end{eqnarray}
The numerical method for solving Eqs. (\ref{rteanh}) and (\ref{rteanj}) is described
by Sudarsky et al. (2000). However, it is still simpler and more straightforward to
employ the ALI-based method descried in \S\,\ref{ali}.


\subsubsection{$\delta$-function reduction of the phase function}

The phase function is typically computed in a set of discrete values of the
scattering angle $\gamma = \gamma_1, \gamma_2, \ldots \gamma_{N\!A}$, with
$\gamma_1=0$ and $\gamma_{N\!A}=\pi$. However, in many cases the phase function
is a very strongly peaked function of $\gamma$, with a peak at $\gamma=0$
(forward scattering). Any simple angular quadrature is inaccurate because
$g(\gamma_1=0)$ may be by several orders of magnitude larger than $g(\gamma_2)$
even for very small values of $\gamma_2$. Describing the phase function
close to the forward-scattering peak with sufficient accuracy would necessitate
to consider a large number of angles, which would render the overall scheme impractical

A more efficient approach was developed in Sudarsky et al. (2005; Appendix), which
splits the phase function into two components. The first one, $g^\prime$, is defined as
$g^\prime(\gamma_1)= g(\gamma_2)$ and $g^\prime(\gamma_i)= g(\gamma_i)$ for
$i>1$; i.e. $g^\prime$ is the original phase function with a forward-scattering peak 
being cut off. The second
part is expressed through the $\delta$-function, so that the modified phase function 
is written as
\begin{equation}
g(\gamma) = g^\prime(\gamma) + \alpha \delta(\gamma),
\end{equation}
where $\alpha$ is determined by a requirement that the modified  phase function is 
normalized to unity, i.e.
\begin{equation}
\frac{1}{2} \int_{-1}^1i g(\xi)\, d\xi 
= \frac{1}{2} \int_{-1}^1 g^\prime(\xi)\,  d\xi + \frac{\alpha}{2} = 1,
\end{equation}
where $\xi=\cos\gamma$.
With this phase function, one can write down the source function (\ref{sfanis}) as
(skipping an indication of the frequency dependence)
\begin{eqnarray}
\label{sfanred}
S(\mu,\phi) &=& \frac{1-\epsilon}{4\pi} \int_{-1}^1\!\!d\mu^\prime\!\!\!
\int_0^{2\pi}\!\!\! d\phi^\prime
I(\mu^\prime,\phi^\prime)\, g(\mu^\prime\!,\phi^\prime\!,\mu,\phi) + \epsilon B \nonumber \\
&=& \frac{1-\epsilon}{4\pi} \int_{-1}^1\!\!d\mu^\prime\!
\int_0^{2\pi}\!\!\! d\phi^\prime
I(\mu^\prime\!,\phi^\prime)\, g^\prime(\mu^\prime\!,\phi^\prime\!,\mu,\phi) 
+ \epsilon B\nonumber \\
&+& (1-\epsilon) \alpha I(\mu,\phi).
\end{eqnarray}
The last term, $(1-\epsilon) \alpha I(\mu,\phi)$, represents a creation of photons with
the rate proportional; to the specific intensity, and therefore acts as a reduction of the 
absorption coefficient and thus the optical depth. This is quite natural because the 
forward scattering reduces the extinction of radiation because a photon removed 
from the beam is immediately added to it, and thus cancels the previous act of
photon absorption.

\subsubsection{Combined moment equation in the presence of anisotropic scattering}

The above formalism applies for the formal solution  of the transfer equation
in the case the thermal structure is given. However, to consider the effects of
anisotropic scattering to determine the atmospheric structure, we need to consider 
an equation
for the mean intensity $J$, analogous to Eq. (\ref{rte}). For simplicity, we consider
a $\phi$-averaged case, but the full $\mu$- and $\phi$-dependent case is
analogous.  

Starting with the transfer equation (\ref{rtean2}) with the source function given
by (\ref{sfan}), the moment equations obtained by integrating over $\mu$, and 
by multiplying by $\mu$ and integrating, are as follows
\begin{equation}
\label{hmoma}
\frac{dH}{d\tau} = J -S = \epsilon(J-B),
\end{equation}
because 
\begin{eqnarray}
\frac{1}{2}\!\int_{-1}^1\!\! d\mu\,\, \frac{1}{2}\!\int_{-1}^1\!\!  d\mu^\prime\, p(\mu^\prime,\mu)
I(\mu^\prime)\ \ \ \ \ \  \nonumber \\
= \frac{1}{2}\!\int_{-1}^1\!\!  d\mu^\prime I(\mu^\prime)\,\,
\frac{1}{2}\!\int_{-1}^1\!\!  d\mu\, p(\mu^\prime,\mu) = J.
\end{eqnarray}
The second moment equation presents more problems because while
$(1/2)\int_{-1}^1 d\mu\, p(\mu^\prime,\mu) =1$, the analogous quantity
$(1/2)\int_{-1}^1 d\mu\, \mu\, p(\mu^\prime,\mu) \not=1$, unless $p$ is
an even function of $\mu$.
One can however introduce a form factor
\begin{equation}
\beta\equiv\frac{1}{J}\!\left[ \frac{1}{2}  \int_{-1}^1\!\! d\mu^\prime\, I(\mu^\prime)
\,\,\frac{1}{2}\! \int_{-1}^1\!\! d\mu\, \mu\, p(\mu^\prime,\mu) \right],
\end{equation}
so that the second moment equation can be written as
\begin{equation}
\frac{dK}{d\tau} = H - (1-\epsilon)\beta J.
\end{equation}
The combined moment equation, using Eq. (\ref{hmoma}) and the 
traditional Eddington factor defined by (\ref{vef}), becomes
\begin{equation}
\label{rtejanis}
\frac{d^2(fJ)}{d\tau^2} = \epsilon(J-B) - \frac{d}{d\tau}[(1-\epsilon)\beta J].
\end{equation}
Analogously to the Eddington factor, the new factor $\beta$ is determined
during the formal solution, and is kept fixed in the next linearization step 
where Eq. (\ref{rtejanis}) is used as one of the basic structural equations. 
The second term on the right-hand side is discretized using a three-point difference
formula, analogously as described in Appendix A. The important
point to realize is that the global tri-diagonal structure of resulting matrices is
preserved, so that the global linearization procedure, e.g. the Rybicki scheme,
is unchanged. The effects of anisotropy are contained in the form factor $\beta$,
and also indirectly in the Eddington factor $f$ which is modified with respect to the
isotropic case.

To the best of our knowledge, the procedure outlined above was not yet used
for actual computations. Studies that examined an importance of anisotropic
scattering on condensates  (e.g., Sudarsky et al. 2005) calculated a formal solution
of the transfer equation for the specific intensity, with the source function 
given by (\ref{sfanis}) or (\ref{sfanred}), but only for a given atmospheric structure
(i.e., the $T$-$P$ profile). They did not iterate to obtain a  modified temperature structure.
These effects are expected to be small, but this remains to be verified using the 
procedure outlined above.


\subsection{Application of the Accelerated lambda iteration}
\label{ali}

We describe here a formalism for the general,
$\mu$- and $\phi$-dependent case; an analogous formalism applies for the
azimuthally-averaged, $\phi$-independent, case.
The transfer equation is written as (suppressing the frequency subscript)
\begin{equation}
\label{rteanis2}
\mu \frac{dI_{\mu\phi}}{d\tau} = I_{\mu\phi} - S_{\mu\phi},
\end{equation}
where the source function is given by Eq. (\ref{sfan}), i.e.,
\begin{equation}
\label{sfan2}
S_{\mu\phi} = (1-\epsilon) a_{\mu\phi} J + \epsilon B,
\end{equation}
with the factor $a_{\mu\phi}$ given by Eq. (\ref{adef}). 
Solution of Eq. (\ref{rteanis2}) can be written as
\begin{equation}
\label{ilambda}
I_{\mu\phi} = \Lambda_{\mu\phi} [S_{\mu\phi}],
\end{equation}
where $\Lambda$ is an operator that acts on the (total) source function to yield
the specific intensity. Although Eq. (\ref{ilambda}) is written in an operator form, 
we stress that the
$\Lambda$-operator does not have to be assembled explicitly; Eq. (\ref{ilambda})
should rather be understood as a {\em process} of obtaining the specific intensity from
the source function.
In fact, a construction of an explicit $\Lambda$ operator (i.e.,
a matrix, upon discretizing) would be possible, but cumbersome and
rather time consuming. It is never done in actual astrophysical applications. 

The basic idea of the Accelerated Lambda Iteration (ALI) class of methods 
is to write Eq. (\ref{ilambda}) as an iterative process,
\begin{equation}
\label{ali1}
I_{\mu\phi}^{\rm new} = \Lambda^\ast_{\mu\phi} [S^{\rm new}_{\mu\phi}] + 
(\Lambda_{\mu\phi} - \Lambda^\ast_{\mu\phi}) \left[S_{\mu\phi}^{\rm old}\right],
\end{equation}
where $\Lambda^\ast_{\mu\phi}$ is a suitably chosen approximate operator.
Equation (\ref{ali1}) is exact at the convergence limit. The ``new'' mean intensity
is given by
\begin{equation}
\label{ali2}
J^{\rm new} = \frac{1}{4\pi}\int_0^{2\pi}\!\!d\phi \int_{-1}^1\! d\mu\, I^{\rm new}_{\mu\phi}.
\end{equation}
Using Eqs. (\ref{ali1}) and (\ref{sfanis}) in (\ref{ali2}), one obtains, after some
algebra [for details, refer to Hubeny \& Mihalas (2014, \S\,13.5)]
\begin{equation}
\label{jfs0}
\delta J \equiv J^{\rm new} - J^{\rm old} =
\left[I - (1-\epsilon)\bar\Lambda^\ast \right]^{-1} \left[J^{\rm FS} - J^{\rm old}\right],
\end{equation}
where $I$ is the unit operator, and
\begin{equation}
\bar\Lambda^\ast = \frac{1}{4\pi}\int_0^{2\pi}\!\!d\phi  \int_{-1}^1\! d\mu\, 
a_{\mu\phi} \Lambda^\ast_{\mu\phi},
\end{equation}
is the angle-averaged approximate operator. Finally,
\begin{equation}
\label{jfs}
J^{\rm FS} = \frac{1}{4\pi} \int_0^{2\pi}\!\!d\phi \int_{-1}^1 d\mu\, \Lambda_{\mu\phi}
[S^{\rm old}_{\mu\phi}] 
\end{equation}
is a newer value of the mean intensity obtained from the formal solution of the
transfer equation with the ``old'' source function.

Although there are several possibilities, 
the most practical choice of the approximate operator is a diagonal (i.e., local)
operator, in which case its action is simply a multiplication by a real
number, which we also denote as $\Lambda^\ast$ (or its angle-averaged
value as $\bar\Lambda^\ast$). The correction to the
mean intensity is then simply
\begin{equation}
\label{jfsa}
\delta J = \frac{J^{\rm FS} - J^{\rm old}}{1-(1-\epsilon)\bar\Lambda^\ast}.
\end{equation}

Before proceeding further, we employ Eq. (\ref{jfs0}) to point out some basic 
properties of the ALI scheme, and to explain a motivation for using it.

If one sets $\Lambda^\ast=0$, one recovers the traditional Lambda iteration, in
which $J^{\rm new} = J^{\rm FS}$, i.e. the iteration procedure simply alternates
between solving the transfer equation with the known source function,
and recalculating the source function with just determined intensity of radiation.
This procedure is known to converge very slowly if the scattering term dominates,
i.e.,  if the single scattering albedo is very close to unity. 

On the other hand, if one sets $\Lambda^\ast = \Lambda$, one recovers an exact solution
which can be done in a single step without a need to iterate.  However, the
inversion of the $\Lambda$ operator (matrix) may be quite
costly. Therefore, in order an ALI scheme to be efficient, $\Lambda^\ast$ must be
chosen in such a way that it is easy and cheap to invert, yet still leads to a fast
convergence of the overall iteration process.

From the physical point of view, we see that the ALI iteration process is driven, 
as is the ordinary Lambda iteration, by the difference between the old source function 
(or mean intensity) and the
newer source function (mean intensity) obtained from the formal solution.
But Eq. (\ref{jfs0}) shows that in the case of  ALI this difference is effectively amplified 
by an \textit{acceleration operator} $[1-(1-\epsilon)\Lambda^\ast]^{-1}$. For example,
any diagonal (i.e.~local) $\Lambda^\ast$ operator must be constructed to satisfy
$\Lambda^\ast(\tau)\rightarrow 1$ for large $\tau$ (because $I_\nu\rightarrow
S_\nu$ for large $\tau$). In a typical case $\epsilon\ll 1$, and thus
$[1-(1-\epsilon)\Lambda^\ast]^{-1} \rightarrow \epsilon^{-1}$, so that the
acceleration operator does in fact act as a large amplification factor. 

From the mathematical point of view, an idea of solving large linear
systems by splitting the system matrix into two parts, one being inverted, and the other one
being used to compute an appropriate correction to the solution,  goes back  to Jacobi in 
the mid nineteenth century. In the current
literature these methods are known as {\em preconditioning} techniques.

A comprehensive review of their mathematical properties that are important in
the context of astrophysical radiative transfer is given in the recent textbook by 
Hubeny \& Mihalas (2014, \S\,13.2). The most important conclusion is that
the convergence speed of any preconditioning method is determined by the largest
eigenvalue of the amplification matrix, which is given through the original matrix and
the preconditioner, in our case by $\Lambda$ and $\Lambda^\ast$.  This gives an
objective criterion for judging the quality of the chosen approximate operator. From this
analysis (first done by Olson et al. 1986) it follows that a diagonal (local) $\Lambda^\ast$,
given as a diagonal part of the exact $\Lambda$, 
provides a reasonable compromise between the convergence speed and a time consumption
per iteration. Its construction, in one particular case, is described below.

Returning back to the present application, 
here is an algorithm for solving Eq. (\ref{rteanis2}) using the ALI method:
\begin{enumerate}
\item  For a given $S^{\rm old}$ (with an initial estimate $S^{\rm old}=B$ or
some other suitable value), perform a formal solution of the transfer equation
fro all directions, but one direction (given $\mu$ and $\phi$) at a time. This yields
new values specific intensity $I_{\mu\phi}$ and also new values of the
angle-dependent approximate operator approximate 
$\Lambda_{\mu\phi}^\ast$ -- see below.
\item By integrating over directions using Eq.\,(\ref{jfs}) obtain new values of
the formal-solution mean intensity $J^{\rm FS}$.
\item Using (\ref{jfsa}), evaluate a new iterate of the mean intensity 
$J^{\rm new}=J^{\rm old} + \delta J$.
\item Update the source function from (\ref{sfan2}) using the newly found mean 
intensity and  repeat steps (i) to (iii) to convergence.
\end{enumerate}


\subsubsection{Construction of the approximate operator}
\label{conlam}
Remaining part of the solution is a construction of the approximate operator $\Lambda^\ast$.
There are several possibilities, depending on which formal solver of the transfer
equation is being used. 

As explained in Hubeny \& Mihalas (2014; \S\,13.3), the matrix elements of the 
$\Lambda$-operator can be formally evaluated by setting the source function 
to the unit pulse function, $S(\tau_{d})=\delta(\tau-\tau_d)$, so that
\begin{equation}
\Lambda_{dd^\prime} = \Lambda_{\tau_d}[\delta(\tau_{d^\prime}-\tau)].
\end{equation}
Therefore, one could obtain the diagonal elements of exact $\Lambda$ by solving 
the transfer equation with the source function given by the $\delta$-function. However, 
in practice one does not have to solve the full transfer equation, but only to collect
coefficients that stand at $S_d$ in the expressions to evaluate $I_d$. 

In the case of DFE scheme, one proceeds along the recurrence relations 
(\ref{dfem}) and (\ref{dfep}) to compute
\begin{eqnarray}
\label{lpm}
L_{d+1}^- &=& b_d/a_d,\\
L_{d}^+ &=& [2(\Delta\tau_{d+1/2} + 1)\, L_{d}^- + b_d]/a_d
\end{eqnarray}
where $a_d$ and $b_d$ are given by (\ref{dfea}) and (\ref{dfeb}). The complete
diagonal element of the (angle-dependent)  elementary operator is obtained, 
in parallel with Eq. (\ref{dfei}), as
\begin{equation}
\label{dfelam}
\Lambda^\ast_d(\mu,\phi) \equiv \Lambda_{dd} = 
\frac{L_d^- \Delta\tau_{d+1/2}+ L_d^+ \Delta\tau_{d-1/2}}
{\Delta\tau_{d+1/2}+  \Delta\tau_{d-1/2}}.
\end{equation}
The values at the boundaries are $\Lambda_{dd}=0$ for $d=1$, and $\Lambda_{dd}=L_d^-$
for $d=N\!D$. An evaluation of the diagonal elements for outward-directed rays is
analogous,
\begin{eqnarray}
\label{lpmm}
L_{d}^- &=& b_d/a_d,\\
L_{d+1}^+ &=& [2(\Delta\tau_{d+1/2} + 1)\, L_{d+1}^- + b_d]/a_d
\end{eqnarray}
As stressed in \S\,\ref{dfe}, a solution of the transfer equation using the DFE method
is performed for one direction at a time, so $L$ and $\Lambda$ in Eqs. (\ref{lpm}) -
(\ref{dfelam}) are evaluated
for given $\mu$ and $\phi$. An angle-averaged approximate operator needed to evaluate the
new iterate of the source function or the mean intensity, as in Eq. (\ref{jfsa}), is then given by
\begin{equation}
\bar\Lambda^\ast_d = \frac{1}{4\pi}\int_0^{2\pi}\!\!d\phi  \int_{-1}^1\! d\mu\, 
\Lambda^\ast_d(\mu,\phi).
\end{equation}

In the case of Feautrier scheme, which is however useful only for isotropic
scattering, one uses a special procedure to evaluate an
elementary $\Lambda^\ast$ suggested by Rybicki \& Hummer (1991), see also Hubeny
\& Mihalas (2014, \S\,13.3).


\section{Details of numerical implementation}
\label{details}

\subsection{Treatment of opacities and the state equation}
\label{opac}

Unlike model stellar atmospheres, where the opacities are evaluated on the fly,
here we use pre-calculated extensive tables of opacity 
as a function of frequency, temperature, and density (or pressure).
Such an approach is used for instance in the computer code
{\sc Cooltlusty} (e.g. Hubeny et al. 2003; Sudarsky et al. 2003),, which is a variant
of the stellar atmosphere code
{\sc tlusty} (Hubeny 1988; Hubeny \& Lanz 1995).


The opacity table can be set either 
(i)  as the total opacity of all gaseous species, or
(ii) opacities of the individual species separately. In the latter case,  the table contains
the corresponding cross sections $\sigma$. This approach is mandatory when
treating departures form chemical equilibrium. On the other hand, one needs an
additional table of concentrations of the species, or an analytical or empirical 
prescription how to evaluate them.

In both cases, the individual values of $\kappa_i(\nu_j)$ or $\sigma_i(\nu_j)$ for
the individual frequencies are set using one of the two possible approaches:
\begin{enumerate}
\item Using the idea of Opacity Sampling (see, e.g. Hubeny \& Mihalas 2014,
\S\,18.5) that
is used  in the stellar atmospheres applications. In the planetary context, it is
known as the {\em line-by-line approach}. It consists simply of evaluating the exact opacity
at the actual set of frequencies $\nu_j$. If the set of frequencies is dense enough, 
this scheme essentially amounts to an exact representation of the opacity.
However, if the frequency points are not spaced sufficiently densely, this approach 
may miss cores of strong lines, or windows between them.
\item Using the idea of Opacity Distribution Functions (ODF), also often used in the
context of stellar atmospheres (e.g. Hubeny \& Mihalas 2014; \S\,17.6 and 18.5). 
This approach consists of three parts:
\begin{description}
\item (a) Dividing the global range of frequencies into a set of relatively narrow intervals
(typically $10^2$ to several times $10^3$ intervals);
\item (b) For each interval, one first computes a detailed line-by-line opacity with a very
high frequency resolution, and then resamples the opacity to form a monotonic
function of frequency, called ODF.
\item (c) This function is represented by a small number (typically of the order of $10^1$)
frequency points.
\end{description}
This approach is analogous to the so-called {\em correlated k-coefficient method} 
(Goody et al. 1989; for an illuminating discussion, see Burrows et al. 1997),
used in the planetary context.
An advantage of this approach is that both high- and low-opacity points
are well represented; however, a disadvantage is that the position of, say, the 
highest peak in the true opacity distribution is generally different from the position 
of the peak of an ODF.
Nevertheless, if the intervals are chosen to be
small, the resulting errors are also small.

In the context of SMO model atmospheres, where the opacity is dominated by
strong molecular bands composed of many closely spaced lines, the ODF approach
is expected to work better than in the stellar atmosphere context where an ODF
represents a set of relatively well separated lines.
\end{enumerate}

From the practical point of view, one needs several tables:\\ [2pt]
-- a table (or a set of tables) of the gaseous opacity;\\ [2pt]
-- a table of the total Rayleigh scattering opacity;\\ [2pt]
-- a set of Mie scattering cross sections for the individual condensates;\\ [2pt]
-- a set of cross sections for absorption of the individual condensates.

The corresponding derivatives with respect to the temperature, needed
to evaluate the Jacobian, are computed numerically.

Analogously, one needs pre-calculated tables of density as a function of
$T$ and $P$ and, for evaluating the thermodynamic parameters needed
for treating convection, the internal energy ($E$) or entropy ($S$)
as a function of $T$ and $P$. Summarizing, one needs two more tables:\\ [2pt]
-- a table of $\rho=\rho(T,P)$;\\ [2pt]
-- a table of $E=E(T,P)$ or $S=S(T,P)$.

In this manner, all calculations that are connected to chemical equilibrium 
and determining the opacities are separated from the calculation of the atmospheric 
structure.


\subsection{Setting up the cloud bases}
\label{cloudbase}

Ideally, the position of the (upper) cloud base should be given as an intersection
of the current $T$-$P$ profile and the condensation curve. The lower cloud base
is an artificial concept. If it is set through the condensation curve of the
surrogate species, or is set at a fixed temperature, it mimics the
situation where there are many condensates with actual condensation curves between
these two limits,
so that the given species is in fact a representative of a cumulative effect of
many condensates.

For instance, Burrows et al (2006) chose forsterite (Mg${}_2$SiO${}_4$) to represent
about 20 individual species of magnesium and aluminum silicates; with upper cloud
base determined through the forsterite condensation curve, and the lower base
at fixed temperature $T=2300$ K, which roughly corresponds to a characteristic
highest condensation temperature of other silicates (see Fig.1 of Burrows et
al., 2006).

This procedure works well if the cloud is located in an optically thick portion of the atmosphere.
However, numerical experience showed that in cases where the upper or lower
base is located in an optically thin part of the atmosphere, tcloud position may
oscillate between two or more locations, and in fact in no location can one
obtain a cloud position fully consistently with the atmospheric structure.
For instance, at certain iteration a cloud base is determined to be at a certain, 
say low-$P$ position. When the cloud is located there, its influence modifies the
temperature, and as a consequence the cloud moves to higher $P$.
Again, this modifies the temperature, and in the next iteration the cloud
moves back to the low-$P$ location. After a few iterations, the model starts
to oscillate between two identical cloud positions. Moreover, regardless
where the cloud position is set empirically, for instance anywhere between
the two positions mentioned above, the resulting temperature structure
that is obtained after such a cloud is taken into account, moves the
cloud away. In such situations, there is no stationary solution
of the problem. To obtain at least an
approximate solution in those cases, several procedures were devised.
They were used by Burrows et al. (2006) and Hubeny \& Burrows (2007), but 
not explicitly described there.

In those procedures, one first calculates the cloud base position that depends
only on the current atmospheric structure. As mentioned above, there are three possibilities:

(1) Setting the cloud base at an intersection of the $T$-$P$ profile 
with the condensation curve -- the ``exact'' way.

(2) Setting the cloud base at a specified temperature (which corresponds to an
approximate condensation curve that is independent of pressure).

(3) Setting the cloud base at a specified pressure. In this case, since the pressure
is unchanged during iterations, the cloud base is also fixed in space. Obviously,
this is not a good physical model, but this approach may be useful for testing,
and for diagnosing problems when the code cannot find the self-consistent cloud bases.
For instance, one may construct a series of models with many fixed cloud
base positions, and to study which position is closest to a consistent one,
that is to the one where the computed $T$-$P$ profile intersects the condensation
curve closest to the position where the cloud base was set.

The cloud bases determined by any of the procedures
(1) or (2) are called ``tentative bases''. 
The tentative cloud bases may be either kept as they are, or may be modified
by several possible procedures:

(a) The position of the new cloud base cannot be moved more that a prescribed number
of depth points.

(b) The actual position of the base is set at the midpoint between the tentative
and the previous base. The ``previous'' base is the final
base determined (by any procedure) at the preceding iteration.

(c) The actual position of the base is set as a weighted geometrical
mean of the tentative and the previous base. In this case, one computes the
geometrical mean of the pressures at the cloud bases. Specifically, say for the
upper base,
\[
P_0^{\rm actual} = (P_0^{\rm tent})^w \times (P_0^{\rm previous})^{1-w}\, ,
\]
where $w$ is a weight for the geometrical mean, typically set to $w=1/2$, 
i.e., as s true geometrical mean.

\begin{figure}
 \includegraphics[width=\columnwidth,height=5cm]{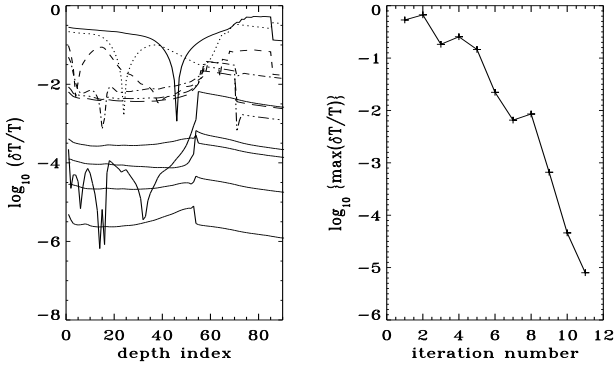}
 \caption{Convergence pattern for a model analogous to that
 displayed in Fig.\,\ref{fig:fii2}, but with adding a forsterite cloud.}
 \label{fig:fic0}
\end{figure}
\begin{figure}
 \includegraphics[width=\columnwidth,height=5cm]{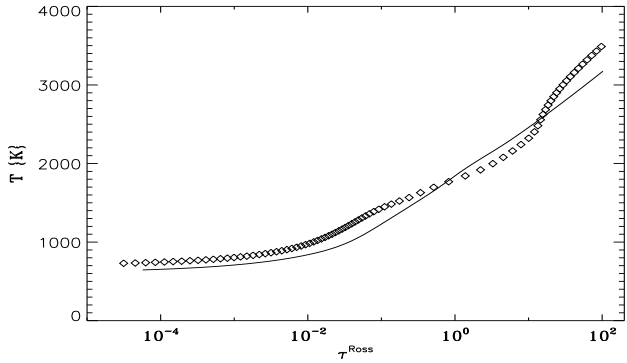}
 \caption{Temperature structure for a model atmosphere with
 $T_{\rm eff}$ = 1500 K, $\log g=5$, computed without clouds (solid line),
 and with a forsterite cloud (diamonds).}
 \label{fig:fic1}
\end{figure}
\begin{figure}
 \includegraphics[width=\columnwidth,height=5cm]{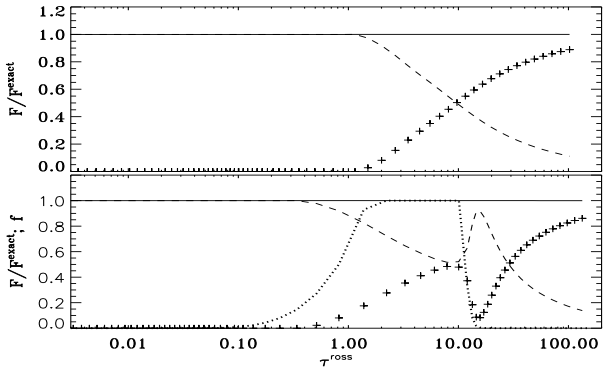}
 \caption{Conservation of the total flux for the model atmospheres displayed in Fig.\,\ref{fig:fic1}.
 Upper panel: model without clouds; lower panel: model with clouds. The basic parameters
 (effective temperature, surface gravity) are otherwise the same
 Here, $F^{\rm exact}\equiv\sigma_{\!R} T_{\rm eff}^4$ is the nominal total flux. 
 Dashed line represents the radiation flux, and crosses represent the convective flux,
 both divided by the total nominal flux.
 Dotted line in the lower panel displays the cloud shape function $f$, which essentially shows
 the position and the opacity distribution of the cloud.}
 \label{fig:fic2}
\end{figure}
\begin{figure}
 \includegraphics[width=\columnwidth,height=5cm]{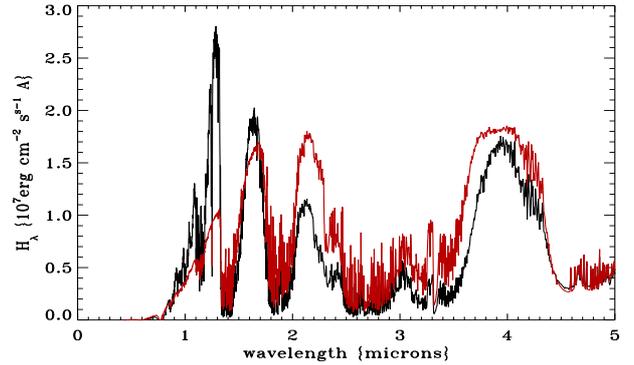}
 \caption{Predicted emergent flux for the models displayed in Fig.\,\ref{fig:fic1}.
 Black line: cloudless model; gray line (red in the online version): cloudy model.}
 \label{fig:fic3}
\end{figure}

Another possible numerical trick is a ``rezoning'' of depth points. It was found
that it is more accurate and numerically mode stable to add several depth points
at the newly determined low-pressure base of the cloud deck and immediately above it.
Otherwise, if there are too few depth points in the region of exponential decline of the
cloud-shape function on the low-pressure side of the main cloud, the opacity of the
cloud would be overestimated. Analogously, if there is no depth point exactly at the cloud
base, the opacity of the cloud is underestimated. 

Some results that illustrate an influence of clouds are shown in Figs.\,\ref{fig:fic0} -- \ref{fig:fic3}. 
We compare a cloudless model considered earlier with $T_{\rm eff}$ = 1500 K,
$\log g = 5$, to an analogous model with an added forsterite
(Mg${}_2$Si${}$O${}_4$) cloud. The low (high-pressure) cloud boundary is set at a fixed
temperature of $T=2300$ K that simulates an effect of a whole set of other magnesium
silicate condensates, as suggested by Burrows et al. (2006). Notice that even if the lower
cloud boundary is specified at a fixed temperature, it is not fixed in the physical space
because the temperature structure varies from iteration to iteration. The upper (low-pressure)
cloud boundary is set exactly at the intersection of the $T-P$ profile and the forsterite condensation
curve. The power-law cloud shape parameters defined by Eq.~(\ref{cloudsh}) are set to 
$c_0 = 2$ and $c_1 =10$. The modal particle size is taken to be 100 microns.

Figure \ref{fig:fic0} displays the convergence pattern of a model with clouds, computed using
the Rybicki scheme. As is clearly seen, the convergence is again quite fast a very stable;
the whole computation took about 30 s on the same MacBook Pro laptop as mentioned 
in \S\,\ref{ryb}.
Figure~\ref{fig:fic1} shows the temperature structure, displayed as the temperature as 
a function of Rosseland optical depth for both, cloudless and cloudy models. Differences
in the temperature structure are clearly seen. 

The effects of the cloud are best seen on
a plot of the total radiative and convective energy flux, displayed in  in Fig.\,\ref{fig:fic2}. 
The upper panel shows the cloudless model, which exhibits a smooth rise of 
$F^{\rm conv}/(\sigma_{\!R} T_{\rm eff}^4)$ toward deep layers, starting around 
$\tau_{\rm ross} \approx 1$. From the numerical point of view, notice 
that the total flux is conserved within about 0.05\%; this is not seen on this plot but is shown
later in Fig.\,\ref{fig:fflu}. The lower panel represents an analogous plot for the cloudy model,
together with the cloud shape function. The later plot clearly shows that the cloud contributes
to the total opacity
at Rosseland optical depths roughly between 1 and 10. Because of an additional opacity as 
compared to the cloudless model, the temperature gradient is flatter in this region, and
consequently the radiative flux is somewhat lower. 
The relative portion of the convective flux in this region thus somewhat increases.
In contrast, in the region just below the cloud, the temperature gradient increases and so does the
radiative flux, and consequently the portion of the convective flux decreases dramatically.

Finally, we show in Fig.\,\ref{fig:fic3} the predicted emergent flux for both models. The
main effect of clouds is to fill the opacity windows at 1.2 and 1.6 microns where the
cloudless model exhibits the highest peaks of the spectral energy distribution. By virtue
of the radiative equilibrium, this energy has to be redistributed in other spectral regions,
and therefore the flux increases essentially everywhere for wavelengths larger than
about 1.8 microns.


\subsection{Global formal solution}
\label{formal}

The term ``global formal solution'' refers to
the set of all calculations between two iterations of the overall iteration
(i.e., linearization) scheme. 

The main part of this procedure is a
solution of the radiative transfer equation for specific intensities
and an evaluation of the Eddington factors, as described above in \S\,\ref{forrte}.

In parallel with, or on top of, this procedure, one performs other
``formal'' solutions, essentially updating one state parameter by solving
the appropriate equation, while keeping other state parameters fixed. 
For instance, and most importantly, one solves the radiative/convective
equilibrium equation to update temperature in the convection zone and below it.
To this end, several procedures were devised for convective models
to iteratively improve the $T$-$P$ profile before entering the next linearization step. 
In most cases, using such procedures has very favorable consequences for the
convergence properties, or even prevents an otherwise  violent divergence of the 
iteration scheme. These procedures will be described next in \S\,\ref{convec}.

For models with clouds, one then determines the new positions of the cloud bases as described in
\S\,\ref{cloudbase}. This changes the opacity as a function of depth, so one has to
perform another formal solution of the radiative transfer equation, as well as the
radiative/convective equilibrium, and the whole procedure may be
iterated several times.


\subsection{Correction of temperature in the convection zone}
\label{convec}

Although the linearization
scheme may in principle converge without additional correction procedures,
in practice it is a rare situation. The essential point is that a
linearization iteration may yield current values of temperature and other
state parameters such that, for instance, the actual logarithmic gradient of temperature in a
previously convective region may spuriously decrease below the adiabatic gradient
at certain depth points. Consequently, these points would be considered as
convectively stable, and in the next iteration the radiative flux would be forced 
to be equal to the total flux. This would lead to a serious
destabilization of the overall scheme, likely ending in a fatal divergence.

It is therefore often necessary to perform certain correction
procedures to assure that the convection zone is not disturbed by spurious
non-convective regions, and analogously the radiative zone is not disturbed
by spurious convective region, 
so that the temperature and other state
parameters are smooth functions of depth before one enters the next
iteration of the overall linearization scheme.
We describe these schemes below.
\subsubsection{Improved definition of convection zone}.
After a completed linearization iteration, one examines the depth points in which
the actual temperature gradient surpasses the adiabatic one. If such a point is solitary,
or if it occurs at much lower pressures than the upper boundary of the convection 
zone in the previous iteration, the point is declared as convectively stable,
and the usual radiative equilibrium equation is solved for it in the
next iteration step. 

On the other hand, if there is/are depth points in which $\nabla < \nabla_{\rm ad}$
(so that they are seemingly convectively stable),
surrounded on both sides by points that are convectively unstable,
$\nabla \geq \nabla_{\rm ad}$, these points are declared as convectively
unstable, and are considered to be part of the convection zone. In such 
a newly defined convection zone, one or both of the following correction procedures
are performed.
\subsubsection{Standard correction procedure}.
The idea of the correction is as follows. In view of eq. (\ref{conv}), the
convective flux is given by
\begin{equation}
\label{conref1}
F_{\rm conv} = F_0 (\nabla-\nabla_{\rm el})^{3/2},
\end{equation}
where 
\begin{equation}
\label{conref2}
F_0 = (gQH_P/32)^{1/2}(\rho c_P T)(\ell/H_P)^2.
\end{equation}
After a completed iteration of the global linearization scheme, one takes
the current values of the state parameters and the radiation flux, and computes,
in the convection zone, the new convective flux corresponding 
to this radiation flux so that the total flux is perfectly conserved,
\begin{equation}
\label{conref3}
F_{\rm conv}^\ast =  F_{\rm tot}- F_{\rm rad},
\end{equation}
where $F_{\rm tot} = \sigma_{\!R} T_{\rm eff}^4$.
If $F_{\rm rad}$ is spuriously larger than $F_{\rm tot}$, then $F_{\rm rad}$
is set to $0.999 F_{\rm tot}$.
The new difference of the temperature gradients 
corresponding to this convective flux is then
\begin{equation}
\label{conref4}
\nabla-\nabla_{\rm el} = (F_{\rm conv}^\ast/F_0)^{2/3},
\end{equation}
which is related to $\nabla-\nabla_{\rm ad}$ through
\begin{equation}
\label{conref5}
\nabla-\nabla_{\rm ad} = (\nabla-\nabla_{\rm el})+B \sqrt{\nabla-\nabla_{\rm el}}.
\end{equation}
where $B$ is given by eq. (\ref{convb}). Both $B$ and $\nabla_{\rm ad}$
are computed using the current values of the state parameters. Equation
(\ref{conref5}) thus yields the new gradient $\nabla$ and, with the pressure being
fixed, the new temperature. With the new temperature, one recalculates the
thermodynamic variables, and iterates the process defined by
equations (\ref{conref2}) - (\ref{conref5}) to convergence.

In solving eq. (\ref{conref5}), one proceeds from the top of the convection zone
to the bottom, because the gradient $\nabla$ is numerically given by 
\begin{equation}
\label{nabla1}
\nabla_d \equiv \nabla_{d-1/2} =
\frac{T_d - T_{d-1}}{P_d-P_{d-1}} \frac{P_d+P_{d-1}}{T_d + T_{d-1}}.
\end{equation}
or by
\begin{equation}
\label{nabla2}
\nabla_d  = \ln (T_d /T_{d-1}) / \ln (P_d /P_{d-1}),
\end{equation}
so in order to  evaluate $T_d$
one needs to know $T_{d-1}$ in the previous depth point.
\subsubsection{Refined correction procedure}
The above procedure is improved by recognizing that the coefficient
$B$ is an explicit function of temperature, so $B$ can be expressed
as $B\equiv\beta T^3$. More importantly, the radiation flux is not kept
fixed, but is written as
\begin{equation}
\label{conref6}
F_{\rm rad} \equiv \alpha T^4 \nabla,
\end{equation}
so that instead of keeping $F_{\rm rad}$ fixed, one first computes
$\alpha$ from (\ref{conref6}) for the current values of $T$ and $\nabla$,
and  rewrites combined equations (\ref{conref3}) --(\ref{conref5}) as a
non-linear equation for temperature,
\begin{equation}
\label{conref7}
\nabla(T) = \nabla_{\rm ad} +
\left(\frac{F_{\rm tot} - \alpha T^4 \nabla(T)}{F_0}\right)^{2/3}
+ \beta T^3 \left(\frac{F_{\rm tot} - \alpha T^4 \nabla(T)}{F_0}\right)^{1/3},
\end{equation}
where the parameters $\alpha$ and $\beta$ are held fixed. Equation (\ref{conref7})
is solved by the Newton-Raphson method, again going from the
top of the convection zone to the bottom. 

These procedures were developed by Hubeny \& Burrows (2007), but not explicitly described
there. Experience showed that they may be very helpful, but should be used judiciously.
The best strategy is to start using them around the third or fourth iteration of
the linearization scheme (otherwise, the radiation flux is so far from the correct
value that the correction cannot work properly), and to stop using them at some
later (e.g., 15th) global iteration. The reason for this cutoff is that an application of the 
refinement procedures for
an almost converged model may lead to an oscillatory behavior of the temperature
corrections, in the sense that
the refinement procedures change the temperature slightly, while the subsequent 
linearization iteration changes it back.


\section{Gray and pseudo-gray models}
\label{graymu}

It is instructive to consider the so-called gray, or pseudo-gray models.
These are approximate models, but they serve two purposes: 
(i) they can be used as initial models for the linearization scheme, and
(ii) they can provide a valuable physical insight into the properties of the
computed atmospheric structure.

They are based on the two moment equations of the transfer equation,
Eqs (\ref{hmom}) and (\ref{kmom}), rewritten to contain derivatives with respect
to the column mass $m$,
and integrated over frequencies, namely
\begin{eqnarray}
\label{hmomg}
\frac{dH}{dm}&=&\kappa_J J - \kappa_B B,\\
\frac{dK}{dm}&=&\chi_{\!H} H,
\end{eqnarray}
where
\begin{equation}
[J,H,K] \equiv \int_0^\infty [J_\nu, H_\nu, K_\nu]\, d\nu
\end{equation}
are the frequency-integrated moments of the specific intensity, and
\begin{eqnarray}
\kappa_J &\equiv& \int_0^\infty\!\! (\kappa_\nu/\rho) J_\nu d\nu /J, \\
\kappa_B &\equiv& \int_0^\infty\!\!  (\kappa_\nu/\rho) B_\nu d\nu /B, \\
\chi_{\!H} &\equiv& \int_0^\infty\!\!  (\chi_\nu/\rho) H_\nu d\nu /H,
\end{eqnarray}
are the absorption mean, the Planck mean, and the flux-mean opacities,
respectively. Here
\begin{equation}
B\equiv\int_0^\infty\!\! B_\nu d\nu = (\sigma_{\!R}/\pi) T^4,
\end{equation}
is the frequency-integrated Planck function, which is proportional to $T^4$.
As is customary, the mean opacities are defined using the
monochromatic opacities per gram.
Notice that $\kappa_J$ and $\kappa_B$ are defined through the true absorption
coefficient (without scattering), while $\chi_{\!H}$ is defined through the total
absorption (extinction) coefficient.

Assuming radiative equilibrium, $dH/dm=0$, Eq. (\ref{hmomg}) reduces to
\begin{equation}
\label{regr1}
\kappa_J J = \kappa_B B, \quad{\rm or}\quad B=(\kappa_J/\kappa_B) J,
\end{equation}
which shows that the temperature structure is given through the ratio
of the absorption mean to the Planck mean opacities, and the integrated
mean intensity, which is given by the solution of the transfer
equation. From the second moment equation we have
\begin{equation}
\label{regr2}
K(\tau_H) = H\tau_H + K(0) = (\sigma_{\!R}/4\pi) T_{\rm eff}^4 \tau_H + K(0),
\end{equation}
where $d\tau_H = \chi_{\!H} dm$ is the optical depth associated with the
flux-mean opacity. We express the moment $K$ through $J$ via an integrated
Eddington factor, $f_K\equiv K/J$, and using an integrated second Eddington factor,
$f_H\equiv H(0)/J(0)$, Eq. (\ref{regr1}) together with (\ref{regr2}) gives
(see also Hubeny et al. 2003)
\begin{equation}
\label{graygen}
T^4 = \frac{\kappa_J}{\kappa_B} \left[ \frac{3}{4} T_{\rm eff}^4 
\left(\frac{1}{3f_K}\tau_H+
\frac{1}{3f_H} \right) + \frac{\pi}{\sigma_{\!R}} H^{\rm ext} \right].
\end{equation}
This expression is {\em exact}, but is only formal because $\kappa_J$,
$f_K$, $f_H$, and $\tau_H$ are not a priori known. However, this expression
is very useful if one makes some additional approximations.
\begin{description}
\item {\it Classical gray model without irradiation}.  It assumes that
the opacity is independent of frequency. In this case one has an exact
mathematical solution,
\begin{equation}
\label{graycl}
T^4 = \frac{3}{4}\, T_{\rm eff}^4\, [\tau+q(\tau)],
\end{equation}
where $q(\tau)$ is the Hopf function, a monotonically varying function
between $q(0) = 1/\sqrt{3} \approx 0.577$ and $q(\infty) \approx 0.71$.
Temperature structure given by (\ref{graycl})
is {\em exact} for a truly frequency-independent (gray) opacity,
but it can be used as a useful starting approximation for any opacity, provided
that $\tau$ is presented by a properly chosen mean opacity. 
As follows from the general expression
(\ref{graygen}), the appropriate opacity should be an approximation of
the flux mean opacity. It turns out that such an approximation is the Rosseland 
mean opacity. Specifically, in the
deep layers where the diffusion approximation applies,
\begin{equation}
H_\nu \approx \frac{1}{3}\frac{dB_\nu}{d\tau_\nu}= 
\frac{1}{3}\frac{dB_\nu}{(\chi_\nu/\rho)dm} =
\frac{1}{3}\frac{1}{(\chi_\nu/\rho)}\frac{dB_\nu}{dT}\frac{dT}{dm},
\end{equation}
and therefore
\begin{equation}
\chi_{\!H} = \frac{\int_0^\infty\!\!  (\chi_\nu/\rho) H_\nu d\nu }
{\int_0^\infty\!\!  H_\nu d\nu }
\approx \frac{\int_0^\infty (dB_\nu/dT)\, d\nu}
{\int_0^\infty [1/(\chi_\nu/\rho)](dB_\nu/dT)\, d\nu} \equiv \chi_{\!R},
\end{equation}
where the second equality is the definition of the Rosseland opacity.
\item {\it Gray model with Eddington approximation}. In our notation, the
Eddington approximation sets $f_K=1/3$ and $f_H=1/2$, and the Hopf function
is taken as constant, $q(\tau)=2/3$. Equation (\ref{graycl}) still applies. 
\item {\it Eddington approximation, but allowing for non-gray opacity}.
In this case, the temperature structure is
\begin{equation}
T^4 = \frac{\kappa_J}{\kappa_B}\left(\frac{3}{4}\,  T_{\rm eff}^4 \big[\tau+2/3\big] \right).
\end{equation}
\item {\it Eddington approximation, with non-gray opacity, and with external irradiation}.
\begin{equation}
T^4 = \frac{\kappa_J}{\kappa_B}\left(\frac{3}{4}\,  T_{\rm eff}^4 \big[\tau_H+2/3\big] + 
W T_\ast^4\right),
\end{equation}
where the external irradiation flux is expressed through the effective temperature
of the irradiating star, $T_\ast$, and the dilution factor, $W$, given by Eq. (\ref{dilw}).
As shown by
Hubeny et al. (2003), this expression helps to understand a possible temperature
rise at the surface of strongly irradiated planets,, and even the fact that under 
certain circumstances one can
obtain two legitimate solutions of the structural equations -- one for the temperature
monotonically decreasing outward, and one exhibiting a temperature rise toward
the surface. 

Mathematically speaking, these effects arise due to an inequality of
the absorption mean and the Planck mean opacities in the surface layers, namely that
$\kappa_J/\kappa_B$ may become significantly larger than unity. 
The reason for this is that the Planck mean
opacity weighs the monochromatic opacity by $B_\nu(T)$, the Planck function 
at the {\em local} temperature,
while $\kappa_J$ close to the surface weighs the monochromatic opacity by 
$B_\nu(T_\ast)$, the Planck function
at the effective temperature of the irradiating star, $T_\ast$, 
which is significantly
larger than $T$. If, in addition, one has a strong opacity source acting in the optical
region (where the stellar irradiation has the maximum), one can easily obtain
$\kappa_J/\kappa_B \gg 1$ close to the surface. Further from the surface, where
less incoming radiation penetrates, $\kappa_J \rightarrow \kappa_B$, which leads
to a decrease of the local $T$ as compared to the surface value.
A more comprehensive discussion is presented in Hubeny et al. (2003) and
Hubeny \& Mihalas (2014; \S\,17.7).
\item {\it Two-step gray models}.
A variant of the above approaches is a two-step gray model, which divides the
whole frequency range into two regions, typically a "visible" and an "infrared", one,
and assumes
a frequency independent opacity $\chi_{\rm vis}$ and $\chi_{I\!R}$, 
with $\chi_{\rm vis} \not= \chi_{I\!R}$, and analogously for $\kappa$ and the
scattering coefficient $s$. In the two regions one typically invokes different
approximations. Such models were developed by Hansen (2008), Guillot (2010)
and Parmentier \& Guillot (2014).
\end{description}

We will not discuss this topic any further because our emphasis here is on 
constructing model atmospheres without any unnecessary approximations.
We use gray or pseudo-gray models just as am initial estimate for subsequent
iterative procedure, or as a pedagogical tool to understand the atmospheric
temperature structure.


\section{Comparison to available modeling approaches and codes}
\label{compar}

Here we briefly describe various modeling approaches and codes used 
in the literature and compare them to the formalism described above.
We stress that we will consider here only the codes and approaches that
aim at determining a {\em self-consistent} atmospheric structure, obtained
by a simultaneous solution of the basic structural equations summarized in
Section 2, or at least a temperature structure that is consistent with the radiation filed.
We will not consider here approaches that employ for instance an ad hoc, or
parametrized, temperature structure and solve just for the radiation field, or using
an approximately described, fixed radiation field to determine the atmospheric structure.

Therefore, in the exoplanet terminology, we will consider here only the {\em forward},
self-consistent codes, but we will not consider the {\em retrieval} codes, such as the
code of Madhusudhan \& Seager (2009, 2011), NEMESIS (Irwin et al. 2008; Barstow et al. 2017),
CHIMERA (Line et al., 2012, 2013), or Tau-REX (Waldmann et al., 2015), to name just a few.

From the basic physical point of view, we will limit ourselves here to hydrostatic,
plane-parallel models, because considering more sophisticated multi-dimensional
dynamical models is a different topic that requires different computational strategies.

\subsection{Philosophy}
\label{phil}

Modeling atmospheres of substellar-mass objects is obviously a young field, whose
beginnings occurred in the mid and late 1990's, shortly after observational discoveries of these
objects. In an endeavor to provide a needed theoretical background, it was deemed
most straightforward to adapt some already available modeling approaches and codes
to the physical
conditions expected to occur in SMO atmospheres. There were two avenues taken
in this regard: (i) adapting modeling codes for stellar atmospheres, and (ii) adapting
codes developed for modeling solar system planets and moons. Both avenues offer
certain advantages and certain challenges, as we will outline below. Only recently, there
appear new codes which were developed from the scratch, and which may potentially offer
a possibility of avoiding drawbacks and biases inherent in adapting existing codes.

We shall briefly discuss the most popular and widely used codes in these three
categories. We stress that this is not meant as a comprehensive review of the subject,
but rather as a brief guide to understand what is involved, from both physical
and numerical point of view, in the present most popular modeling codes.

\subsection{Adapting stellar atmosphere codes}
\label{compar_atmos}

The first category of codes are those  that were created by adapting a code for
computing model stellar atmospheres. It should be pointed out that computing
model stellar atmospheres is a very mature subject, having been developing during 
the last almost seven decades. Even the state-of-the-art NLTE metal-line blanketed models
are around for over two decades. The stakes in the stellar atmospheres theory are also
very high thanks to an unprecedented quality and quantity of high-resolution, high
signal-to-noise spectroscopic observations that put heavy demands of the accuracy
and reliability of theoretical analysis tools.

It is therefore quite natural to model atmospheres of SMOs by adapting existing stellar
atmospheres codes. There are specific features that make computing SMO model atmospheres
easier that computing model stellar atmospheres, and vice versa. We will briefly summarize them
below.

The features that make the SMO models easier to compute are:
\begin{enumerate}
\item In stellar atmospheres, in particular for hot stars, the hydrostatic equilibrium 
equation contains a contribution of radiation pressure,
which involves an additional coupling of the gas pressure (and therefore the mass density)
to the radiation field.

\item For both types of objects,  the opacity varies rapidly with frequency. 
However, for stars, the (mostly) atomic lines are distributed randomly in frequency,
while for SMOs, the (mostly) molecular lines tend to be organized in bands, which
makes it more suitable to employ various statistical techniques such as the opacity
distribution functions, or, as they are called in the planetary community, the
correlated $k$-coefficients. Also, for stars, there are no frequency regions that
can be treated as purely (or mostly) scattering or purely (or mostly) absorbing.

\item These two issues play a role already in LTE models. For NLTE models,
a major difficulty comes from the fact that the opacities and emissivities
depend on the populations  of levels involved in
the corresponding atomic transitions, which
in turn depend on the radiation field via the kinetic equilibrium equation. The opacities
thus cannot be evaluated a priori as functions of temperature and density, but
have to be computed self-consistently with all the structural equations. There are
typically thousands to tens of thousands atomic energy levels involved in the atomic
transition (lines or continua) that make a significant contribution
to the total opacity. 
Although in the field of SMO model atmosphere, there are studies that consider NLTE
effects (e.g., Fortney et al. 2004), stellar atmosphere models consider NLTE on much
larger scale. For instance, in a grid of model atmospheres of B stars (Lanz \& Hubeny 2007),
one considers about 1130 energy levels and about 39,000 lines of light elements,
and 500,000 to 2 million lines dynamically selected from a list of about 5.6 million lines
of the iron peak elements, in full NLTE.  
\end{enumerate}

All these complications are absent or alleviated for models of SMO's.
Modifying a modern NLTE stellar atmosphere code thus
mostly involves removing many routines dealing with special issues of
NLTE (an evaluation of transition rates, solving the kinetic equilibrium equation, etc.),
and evaluating opacities and emissivities on the fly, because in any LTE
model atmosphere code, including that for SMO's, it is much more efficient
to use pre-calculated opacity tables.

On the other hand,  computing SMO model atmospheres is more difficult than
computing model stellar atmosphere, particularly for hot stars. We stress that
at the cool end of the main sequence, K and M stars, one meets most of the
challenges listed below for SMO's.
\begin{enumerate}
\item One has to include a solution of chemical networks to 
determine the concentrations of the individual molecular species as functions 
of temperature and pressure. However, this is not difficult numerically or 
algorithmically; the difficulty is mostly in finding appropriate molecular data. 
In any case, this can be done independently of a model construction.
\item As pointed out above, more sophisticated models  needs to consider
departures from chemical equilibrium.
\item  One has to add a treatment of cloud formation, together with an
evaluation of cloud absorption and scattering. This is perhaps the most difficult
part of the process of adapting approaches and codes designed for hotter objects,
because it involves basic physical problems (e.g., determining consistent particle sizes,
their distribution, and  a position of a cloud in the atmospheres), as well as
algorithmic and numerical problems in incorporating these effect in a
self-consistent manner.
\item Although not as serious as other problems listed above, the presence of
strong (and generally anisotropic) external irradiation brings challenges on
adopted numerical schemes, in particular for self-consistent models.
\end{enumerate}

Here is a list of the codes that were created by adapting their stellar 
atmospheric counterpart.

\subsubsection{CoolTLUSTY}
This code is a variant of a general stellar atmosphere (and accretion disk) code
{\sc tlusty}, originally described in Hubeny (1988) and Hubeny \& Lanz (1995).
Its modification for SMO atmospheres, called {\sc CoolTlusty} was briefly described 
in Sudarsky et al. (2003) and Hubeny et al. (2003).

The present paper in fact describes in more detail the physical and numerical
background of {\sc CoolTlusty}. The input atomic and molecular physics and chemistry
is quite flexible. It can either use opacity tables generated using the Burrows \& Sharp (1999)
and Sharp \& Burrows (2007) approach, or any other opacity tables, both for the total opacity,
as well as a set of tables for individual species.
The input properties of condensates (cloud absorption and scattering) can accept any
tables generated by a Mie code. Originally, it was using tables generated as described in
Sudarsky et al. (2000); recently it switched to tables generated by Budaj et al. (2014).

\subsubsection{PHOENIX}
Code PHOENIX was
developed for stellar or even supernova applications, see Hauschildt \& Baron (1999).
The first application for extrasolar giant planets was done by Barman et al. (2001). 
The input physics is analogous to that used in {\sc CooTtlusty}, described above.
The basic difference is the adopted numerical scheme; PHOENIX is using a different 
flavor of the ALI method. It also uses a different set of chemical/molecular data, and
a different treatment fo clouds.

\subsubsection{UMA}
UMA stands for Upsalla Model Atmospheres code (Gustafsson et al. 1974),
somewhat modified by Vaz \& Nordlund (1985). It was further adapted to studies
of extrasolar giant planets by Seager \& Sasselow (1998),
see also Seager \& Sasselow (2000), and Seager et al. (2000). It does not use an
ALI scheme; it solves the radiative transfer equation by the Feautrier method, and
determines the temperature structure self-consistently with the radiation field
by a classical temperature correction.

\subsection{Adapting planetary atmosphere codes}
\label{compar_plan}

Generally, the codes of this category are directly based on approaches
used originally for atmospheres of the solar-system planets or moons.  
Some, but not all, are based on, or use the spirit of, approaches used 
originally for the Earth atmosphere. After the observational detections
of brown dwarfs and extrasolar giant planets in the mid and late 1990's and
early 2000's, some of these codes were adapted to these objects.

In the Earth atmosphere there is a clear distinction between the two 
following wavelength regions:
\begin{enumerate}
\item The optical 
wavelength region (often called ``solar frequencies''), which is optically thin
in most of the visible wavelengths, 
and the transport of radiation is dominated by the scattering processes; and 

\item The infrared region, where the radiation transport is dominated by absorption and
thermal emission.  It should be noted that the atmosphere is opaque in the
short-wavelength regions (UV and X-ray), but these regions are inconsequential
for constructing structural models.
\end{enumerate}

The original Earth-atmosphere codes used that distinction
explicitly to develop suitable approximations of  the radiative transfer equation 
that differ in the optical and the infrared region. The early codes for modeling
solar-system planets often used at least some aspects of this distinction.
However, when applying
such a dichotomous model to significantly hotter or otherwise quite different
conditions in the exoplanets and brown dwarfs, these procedures may become 
less accurate or less efficient than those based on the formalism outlined above.

While the existing codes of this category do still yield valuable results, the
above considerations should be kept in mind when developing new codes for modeling
atmospheres of extrasolar planets of brown dwarfs. Figuratively speaking, it seems
more efficient to treat exoplanets and brown dwarfs as small and cool stars rather 
than hot and big Earths' or solar system planets.

\subsubsection{McKay-Marley code}

The code was first developed by McKay et al. (1989) for calculating atmospheric 
structure and spectra of Titan, and subsequently extended and 
applied for atmospheres of brown dwarfs  by Marley et al. (1996); Burrows et al. (1997),
to the solar-system giant planets by Marley \& McKay (1999), 
and applied for atmospheres of exoplanets
by Marley et al. (1999),  Fortney et al. (2005, 2008), and  subsequently in a large 
number of SMO studies.

Here we list the main assumptions and approaches used by the code, stressing
the differences form the approach described in this paper and/or used in the above
mentioned codes. 

The code determines the $T$-$P$ profile in the following way: In the convection
zone (or possibly multiple zones) the temperature gradient is assumed to be strictly
adiabatic, and all the flux is transported solely by convection. In the radiative zone, where
the strict radiative equilibrium applies, one employs a special temperature-correction 
procedure, which somewhat resembles the Rybicki scheme described above, in the sense
that one forms a vector of the local temperatures, 
${\bf T} \equiv \{T_1, \ldots, T_{N\!R}\}$, where $N\!R$ is the number of depth points in the
radiative zone, and
computes a correction $\delta{\bf T}$ by using the following matrix equation (in our notation)
\begin{equation}
\label{tcormar}
{\bf A}\, \delta{\bf T} = \sigma_{\!R} T_{\rm eff}^4 - {\bf F}({\bf T}_0),
\end{equation}
where ${\bf F}({\bf T}_0)$ is a vector of the total radiative flux in all the depth points 
of the radiative zone, computed for the current vector of temperatures, ${\bf T}_0$..
Equation (\ref{tcormar}) in fact represents a linearization, or a Newton-Raphson solution,
of a non-linear implicit relation between the radiative flux and the temperature,
${\bf F}({\bf T}) = \sigma_{\!R} T_{\rm eff}^4$, expressing the constancy of the total
radiative flux.
Matrix ${\bf A}$ is the corresponding Jacoby matrix, 
$A_{ij} = \partial F_i/\partial T_j$; that is, the $ij$-component of ${\bf A}$
expresses the response of the total flux at depth $i$ to the temperature at depth $j$.
Unlike the Rybicki scheme, the elements of the Jacoby matrix are not evaluated
analytically. Instead, they are obtained by solving a set of additional radiative transfer
equations, by consecutively modifying a single component of vector ${\bf T}$, for instance
$T_j \rightarrow T_j + \Delta T$ (with $\Delta T$ having a small, arbitrary value such as 1 K),
while keeping the other components unchanged, to 
obtain a perturbed flux at all depth points, ${\bf F}^{p,j}$. The elements of the Jacoby matrix 
are then set to 
\begin{equation}
A_{ij} = (F_i^{p,j} - F_i)/\Delta T.
\end{equation}

Radiative transfer equation is solved by a variant of the two-stream approximation, called
two-stream source function method (Toon et al. 1989). It considers an atmosphere 
composed of a set of zones, and assumes that the thermal source function (i.e., the Planck 
function) is a linear function of optical depth within a given zone. The method
essentially solves the first 
moment equation of the radiative transfer equation directly for the radiative flux,  where
some empirical relation between the zero-order moment (mean intensity) and the 
first-order moment (flux) is invoked. This scheme improves the traditional two-stream 
methods in situations where scattering is present, by considering the scattering source
function computed using the proper phase functions, but using the specific intensities obtained
from the  traditional two-stream approximation for the thermal radiation.

The line opacity is treated using a variant of the Opacity Distribution Function approach 
(see \S\,\ref{opac}), 
called here the $k$-coefficient method. The opacity is assumed to be constant within a given depth
zone, which allows one to introduce a $k$-coefficient not as a true opacity distribution 
function, as is done in the stellar context, but directly as a distribution of the transmission 
coefficients.  

In conclusion, 
the adopted method for solving the transfer equation is inherently approximate and
only first-order accurate, in contrast to the Feautrier scheme or DFE 
used in the above approaches, which are second-order accurate (i.e., a numerical
solution of the transfer equation is exact for a piecewise parabolic source function).
However, this is usually not a big concern or a source of inaccuracies of the resulting model.

A potentially more serious source of inaccuracies lies in the treatment of radiative
equilibrium. While the temperature correction expressed by Eq. (\ref{tcormar})
correctly takes into account the fact that a local flux is determined by the 
{\em global} temperature structure,
an evaluation of the elements of the Jacobian numerically by differencing two numerical 
solutions, moreover approximate ones, of the transfer equation, may lead to inaccuracies,
in particular in optically thin regions.

Even more seriously, 
the radiative equilibrium constraint is applied solely
for the flux, and only the condition
$\int\! F_\nu d\nu = {\rm const}$ is checked. A fulfillment of this condition
is viewed as a verification that a model is well converged for the $T$-$P$ profile. 
However, experience gained
from constructing model stellar atmospheres revealed that at the upper, optically thin
portion of the atmosphere, the radiation flux is quite insensitive to the local temperature,
because it is essentially fixed by the source function at the monochromatic optical 
depth around 2/3. The temperature structure in the upper layers may thus remain 
quite inaccurate even if the total flux is conserved
within, say, 1\% or even less. As discussed above, what is needed in upper layers is to
employ the integral form of the radiative equilibrium, 
$\int\kappa_\nu (B_\nu - J_\nu) d\nu = 0$,
which does not seem to be done in this approach.

\begin{figure}
 \includegraphics[width=7cm]{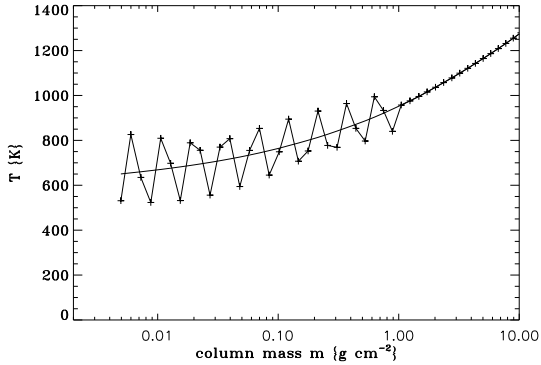}
 \caption{Temperature structure for the radiative zone of a brown dwarf model
 with with $T_{\rm eff}= 1500$ K, $\log g =5$, considered in \S\,\ref{ryb} -- full line,
 and an artificially  perturbed model -- crosses.}
 \label{fig:ftem}
\end{figure}

\begin{figure}
 \includegraphics[width=\columnwidth]{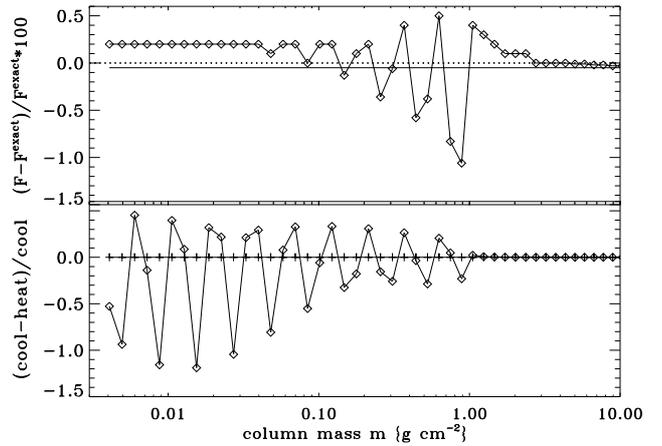}
 \caption{Upper panel: relative difference (in per cents) of the computed 
 radiative and the nominal flux, $\sigma_{\!R} T_{\rm eff}^4$ for the models displayed 
 in Fig.~\ref{fig:ftem}. Full line
 represents the original model, while the diamonds represent the perturbed model.
 Dotted line corresponds to the exact flux with the relative difference equal to zero.
 Lower panel: the net relative cooling rate for the same models. 
 Here the crosses represent the exact model. Notice that while the total
 radiative flux for the perturbed model as still accurate within about 1\%, the net relative
 cooling exhibits huge differences from the exact model, reaching about 120\%.}
 \label{fig:fflu}
\end{figure}

To demonstrate these considerations numerically, we take a brown dwarf model
with $T_{\rm eff}= 1500$ K, $\log g =5$, considered in \S\,\ref{ryb}, and perturb
artificially the temperature structure in the upper layers by adding a damped wavy pattern
with an amplitude $0.3$ times the actual temperature -- see Fig.\,\ref{fig:ftem}. For this model
we recompute the radiative flux, and the heating/cooling rates.
Figure \ref{fig:fflu} shows the flux and the heating/cooling rates. While the computed
radiation flux differs at most by 1\% (close to the column mass $m\approx1$ g$\;$cm${}^{_2}$),
and therefore such model could have easily been declared as reasonably converged,
the net cooling rate, $\int\kappa_\nu (B_\nu - J_\nu)\, d\nu \Big/ \int \kappa_\nu B_\nu d\nu$
shows significant differences from zero. This illustrates the above stated warning that in 
order to assess an accuracy of the model, one needs to check not only a conservation
of the total flux, but also an equality of the heating and cooling rates as stipulated by
the constraint of the radiative equilibrium.

However, we stress that while the above analysis demonstrates that  the McKay-Marley
temperature correction scheme {\em may} lead to an inaccurate determination of the
temperature in the upper layers of an atmosphere, it did not prove that the results
{\em are} necessarily inaccurate. Moreover, even if inaccuracies occur, they are likely
limited to the optically layers, which in turn have relatively little influence on the predicted
emergent radiation.

\subsubsection{Goukenleuque et al.'s code}

Goukenleuque et al. (2000) presented one of the first self-consistent
model atmospheres of an extrasolar giant planet, 51 Peg b in this case. To our
knowledge, this code was not used very much after this study. It takes into account
cloud opacity and scattering, but on the other hand completely neglects convection,
which represents a significant drawback. Radiative transfer equation is solved
approximately, using a variant of the two-stream method with Eddington approximation. 
The code iterates between solving the transfer equation, and subsequently correcting
temperature by solving the radiative equilibrium equation. 

One invokes two nested iteration loops. 
In the inner loop one holds the chemical composition, cloud position, and the opacities
fixed at the current values, and determines the temperature that gives the correct
total flux.
The outer loop takes the $T$-$P$ profile determined in the inner loop,
and computes new chemical equilibrium composition and new opacities
corresponding to this $T$-$P$ profile.
The authors mention that some 1000 (!) iterations were needed in
the inner loop, which, when compared to the linearization scheme outlined
above that requires some 10 - 20 iterations, clearly demonstrates a relative 
inefficiency of this and other similar schemes that do not solve all the 
structural equations simultaneously.

\subsection{Independent, newly developed codes}

\subsubsection{PETIT}
The code is described in detail by Molli\`ere et al. (2015). Although we list
the code as newly developed, the radiative transfer solver and the method
of the solution of the radiative equilibrium equation were  developed already by
Dullemond et al. (2002), and used in a code for computing vertical structure of massive
circumstellar disks. 

Code PETIT solves the radiative
equilibrium and chemical equilibrium equations together with the radiative transfer
equation using a specific application of the variable Eddington factor technique.
Molecular line opacity is treated using the correlated $k$-coefficient method.
The radiative equilibrium equation is considered in a form analogous to our
Eq. (\ref{graygen}), where the Planck mean and the absorption mean opacities,
together with the Eddington factors, are determined iteratively by solving the 
radiative transfer equation frequency by frequency. 
In the convectively unstable layers, the temperature gradient is taken to be adiabatic,
and the integrated mean intensity of radiation is taken as a scaled integrated Planck
function. External irradiation is treated by a variant of the two-stream approximation.

Other approximation is that the PETIT code neglects any scattering process in 
the transfer equation
(see Appendix C1 of Molli\`ere et al. 2015). Also, although the chemical equilibrium
calculations contain some condensed species, cloud formation and opacity is
not considered, which limits the general applicability of the code.

\subsubsection{GENESIS}

The code, together with
its first actual applications, is described in detail in Gandhi \& Madhusuhan (2017).
It essentially uses the structural equations and the numerical procedures
described in this paper, namely the linearization method with the Rybicki
reorganization scheme to solve the coupled radiative transfer together with the
radiative/convective equilibrium equation, and the Feautrier method for
the formal solution of the transfer equation.
Convection is treated using the mixing-length formalism, analogously as described
here. In the present version, the code does not consider cloud opacity and scattering.

\subsubsection{HELIOS}

The code  and its benchmark tests are described in a recent paper (Malik et al. 2017)
Although the code is newly developed from the scratch, it keeps using
approximate and thus potentially inaccurate approaches and numerical schemes,
having their origin in an old Earth/planetary-type 
philosophy of atmospheric modeling, briefly discussed above.  Here is a list
of some shortcomings of the adopted procedure:
\begin{enumerate}
\item The radiative transfer equation is solved by a variant of the two-stream approximation which
uses an analytic solution for the individual layers, assuming either isothermal structure
inside a layer, or a linearly varying Planck function within a layer. The latter still 
yields only a first-order accurate numerical scheme. Although a solution for one 
layer is obtained analytically, the final solution of the transfer 
equation for all layers still requires a numerical procedure. Relative complexity of the
proposed algorithm, which is still approximate, contrasts with the procedure outlined
above which yields an ``exact'' numerical solution, for physical problems of varying 
complexity, in a very simple and transparent way.

\item From the paper (Malik et al. 2017) it appears that the scheme does not include 
convection at all. If this is indeed so, it is a significant drawback
which seriously limits an applicability of the code.

\item Analogously, the published description does not contain any mention of the cloud
opacity and scattering. Such a limitation is however present in other codes mentioned here.

\item In any case, regardless of the deficiencies expressed in (ii) and (iii), the radiative 
equilibrium constraint is treated  as a some sort of time-dependent 
approach to equilibrium. While this is in principle acceptable, the whole procedure still
represents an iterative scheme alternating between (an approximate) solution of the
transfer equation with fixed temperature and a solution (again approximate) of the
radiative equilibrium equation.  Experience gained from computing model stellar
atmospheres revealed that this procedure may converge very slowly, or may even
suffer from the problems of false convergence (i.e., relative changes may become
small, but the current solution is still far from the correct one -- see, e.g. Hubeny \&
Mihalas (2014; \S\,13.2). Furthermore, their formulation of the radiative 
equilibrium equation uses thermodynamic parameters such as specific heat $c_P$, 
and thus ignores the microphysics of the interaction 
of radiation and matter, as contained e.g. in Eq. (\ref{re1}). 
\end{enumerate}


\section{Conclusions}

The aim of this paper was to summarize current physical, mathematical, and
numerical methodology for computing model atmospheres of substellar mass 
objects within a framework of plane-parallel, static models. These two basic 
assumptions make the problem tractable on present-day computers. The remaining
uncertainties and problems are not of an algorithmical or computational  nature, but rather
are caused by the lack of data from other branches of physics and chemistry -- in particular,
data for molecular lines, details of line broadening, formation and detailed properties
of condensed particles, and the rates of chemical reactions for treating non-equilibrium
chemistry, to name just few of the most pressing problems.

Our basic philosophy is the following. While we acknowledge the existence of
many problems and uncertainties that plague our description of the SMO atmospheres,
we feel that the physical formulation and corresponding mathematical treatment
of phenomena that are currently well understood has to be done accurately, reliably,
and without unnecessary approximations and simplifications.

For instance, a treatment of an interaction of radiation and matter, moreover in a highly
non-equilibrium conditions, has been developed to a high degree of sophistication
in stellar astrophysics; for a recent summary, see, e.g, Hubeny \& Mihalas (2014). 
Also, many efficient and fast numerical
algorithms were developed in the last two decades. Yet, many approaches and
numerical codes used for modeling SMO atmospheres are still unnecessarily based
on old and outdated methodologies.
In our opinion, this is caused, at least in part, by the lack of proper communication between
researchers in the fields of planetary and stellar atmospheres. Another reason is the fact that
in the present period of a rapid development of the field of exoplanets and brown
dwarfs, most of the research emphasis is obviously devoted to 
observational issues, like discovering and classifying new objects. 
Even in the subfield of computing SMO model atmospheres most
emphasis if given to applications rather than to a development of new approaches
or to adapting algorithms from different fields.

We have therefore formulated a physical and numerical framework which we believe
should be a standard for dealing with the ``classical'' problem, that is a plane-parallel,
horizontally homogeneous (i.e. 1-D) atmosphere, in the hydrostatic, radiative/convective,
and chemical equilibrium (or with some simple departures from the latter). 
We have stressed that since the radiation field is an important, or even
crucial, ingredient of the energy balance, radiation transport must be treated
accurately, and self-consistently with the global atmospheric structure. 

We believe that this effort does
not represent an imbalanced emphasis on radiation while making serious
approximations for other phenomena, 
for instance the cloud formation. A sophisticated and accurate treatment of an
interaction of radiation and matter is now quite routine, and even not very costly from 
the point of view of computational resources. It is therefore unnecessary or even 
counterproductive to keep applying inefficient and approximate methods for treating 
radiation transfer with the argument that there are many uncertainties in describing the SMO 
atmospheres anyway.

Finally, 
it should be kept in mind that any information, not only about the physical state of a studied
object, but also about a realism of our description, comes only through observed radiation.
Therefore, interpreting spectroscopic observations using unsatisfactory or oversimplified
treatments of radiation may easily yield incorrect results and conclusions. This can be
avoided by using proper methods for treating radiative transfer, for instance those
outlined in this paper, or their future improvements.


\section*{Acknowledgements}
I gratefully acknowledge  the support from the Sackler Distinguished Visitor 
program of the Institute of Astronomy at the University of Cambridge, where most of
the work on this paper was done. My special thanks go to Nikku Madhusudan.  I also 
thank Mark  Marley, Jano Budaj, Ryan Macdonald,  and anonymous
referee for helpful comments to the paper.

\appendix

\section{Discretization and linearization of the basic structural equations}
\subsection{Discretization}
\subsubsection{Radiative transfer equation}
We assume the source function in the form (i.e., for LTE and isotropic scattering)
\begin{equation}
S_\nu= \frac{\kappa_\nu}{\chi_\nu}B_\nu + \frac{s_\nu}{\chi_\nu} J_\nu \equiv
\epsilon_\nu B_\nu + (1-\epsilon_\nu) J_\nu.
\end{equation}
Denoting $d$ the depth index and $i$ the frequency index,
the transfer equation (\ref{rte}), together with boundary conditions 
(\ref{rte_ubc}) and (\ref{rte_lbc2}),
is discretized as follows:
\begin{description}
\item
For $d=1$, the upper boundary condition,
\begin{equation}
\frac{f_{2i}J_{2i}-f_{1i}J_{1i}} { \Delta\tau_{3/2,i}} =
g_i J_{1,i} - H_i^\mathrm{ext} + \frac {\Delta\tau_{3/2,i}}{2} 
\epsilon_{1i} (J_{1i} - B_{1i}),
\end{equation}
where we used the the second-order form of the boundary condition
(Hubeny \& Mihalas, 2014, Eq. 12.50).
\item For $d=2,\ldots,N\!D-1$,
\begin{eqnarray}
\label{discr}
\frac{f_{d-1,i}}{\Delta\tau_{d-1/2,i}\Delta\tau_{di}}\,J_{d-1,i} -
\frac{f_{di}}{\Delta\tau_{di}}\left(\frac{1}{\Delta\tau_{d-1/2,i}}+
               \frac{1}{\Delta\tau_{d+1/2,i}}\right) J_{di} \nonumber \\
 +\frac{f_{d+1,i}}{\Delta\tau_{d+1/2,i}\Delta\tau_{di}}J_{d+1,i} = 
\epsilon_{di}(J_{di} - B_{di})\ \ \ \ \ \ \ \ \ \ \ \ \ \ \ \ \ \ .
\end{eqnarray}
\item For $d = N\!D$, the lower boundary condition,
\begin{eqnarray}
\frac{f_{di}J_{di}-f_{d-1,i}J_{d-1,i} } {\Delta\tau_{d-1/2,i}} &=&
\frac{1}{2}(B_{di}-J_{di}) 
+\frac{1}{3}\frac{B_{di}-B_{d-1,i}}{\Delta\tau_{d-1/2,i}}  \nonumber \\
&-& \frac{\Delta\tau_{d-1/2,i}}{2}
\epsilon_{di}(J_{di} - B_{di}),
\end{eqnarray}
where we again used the second-order form.
\end{description}
In the above expressions
\begin{equation}
\Delta\tau_{d\pm 1/2,i}\equiv
(\omega_{d\pm 1,i}+\omega_{di}) |m_{d\pm 1}-m_d|/2,
\end{equation}
with $\omega_{di}\equiv \chi_{di}/\rho_d$,  and
\begin{equation}
\Delta\tau_{di}\equiv
(\Delta\tau_{d-1/2,i}+\Delta\tau_{d+1/2,i})/2.
\end{equation}
\subsubsection{Radiative/convective equilibrium equation}
Analogously, discretizing the radiative equilibrium equation, one obtains
\begin{eqnarray}
\label{redis}
&\alpha_d&\!\!\!\! \sum_{i=1}^{N\!F} w_i (\kappa_{di}J_{di}-\eta_{di})+ \\ \nonumber
&\beta_{di}&\!\!\!\! \left[ \sum_{i=1}^{N\!F}w_i \frac{f_{dI} J_{di}-f_{d-1,i} J_{d-1,i}}
{\Delta\tau_{d-1/2}} - \frac{\sigma_{\!R}}{4\pi} T_{\rm eff}^4 \right] = 0.
\end{eqnarray}
In the convectively unstable regions, Eq. (\ref{redis}) is modified to read
\begin{eqnarray}
\alpha_d\! \left[
\sum_{i=1}^{N\!F} w_i (\kappa_{di}J_{di}-\eta_{di}) + 
\frac{\rho_d(F_{{\rm conv},d+1/2}-F_{{\rm conv},d-1/2})}{4\pi\Delta m_d}\right]+\quad\quad\quad\,
\nonumber \\
\beta_{di}\!\! \left[ \sum_{i=1}^{N\!F}w_i \frac{f_{dI} J_{di}-f_{d-1,i} J_{d-1,i}}
{\Delta\tau_{d-1/2}} + \frac{F_{{\rm conv},d-1/2}}{4\pi} 
- \frac{\sigma_{\!R}}{4\pi} T_{\rm eff}^4 \right] = 0. \quad\quad\quad \nonumber
\end{eqnarray}
where $\Delta m_d \equiv \Delta m_{d+1/2} + \Delta m_{d-1/2} = (m_{d+1}-m_{d-1})/2$.

\subsection{Outline of the linearization}
The expressions for matrix elements of the Jacobi matrix are straightforward, but tedious
to compute. We just present an example of linearizing Eq. (\ref{discr}).
Let us write this equation as $P_{di}({\bf\psi}) = 0$, which represents the discretized
transfer equation for the frequency point $i$ at depth point $d$. Then
\begin{eqnarray}
(A_d)_{ij}\!\! &\equiv&\!\! - \frac{\partial P_{di}}{\partial J_{d-1,j}} =
\frac{f_{d-1,i}}{\Delta\tau_{d-1/2,i}\Delta\tau_{di}} \delta_{ij},\ \ \ \ \ \ \ \ \ \ \ \ \ \ \  \\
(C_d)_{ij}\!\! &\equiv&\!\! - \frac{\partial P_{di}}{\partial J_{d+1,j}} =
\frac{f_{d+1,i}}{\Delta\tau_{d+1/2,i}\Delta\tau_{di}} \delta_{ij}, \\
(B_d)_{ij}\!\! &\equiv&\!\! \frac{\partial P_{di}}{\partial J_{dj}} =
\left[ \frac{f_{di}}{\Delta\tau_{d,i}}\left(\frac{1}{\Delta\tau_{d-1/2,i}} +
\frac{1}{\Delta\tau_{d+1/2,i}} \right)
+ \epsilon_{di} \right] \delta_{ij} \nonumber \\
\end{eqnarray}
where $d=2,\ldots,N\!D-1$ and $i=1,\ldots,N\!F$.
The columns corresponding to the temperature are
\begin{eqnarray}
(A_d)_{ik} &\equiv& - \frac{\partial P_{di}}{\partial T_{d-1}} =
a_{di} \frac{\partial\omega_{d-1,i}}{\partial T_{d-1}}, \\
(C_d)_{ik} &\equiv& - \frac{\partial P_{di}}{\partial T_{d+1}} =
c_{di} \frac{\partial\omega_{d+1,i}}{\partial T_{d+1}},  \\
(B_d)_{ik} &\equiv& - \frac{\partial P_{di}}{\partial T_{d}} =
- (a_{di}+c_{di}) \frac{\partial\omega_{d,i}}{\partial T_{d}}  \nonumber \\
&+& \frac{\partial\epsilon_{d,i}}{\partial T_{d}} (J_{di}- B_{di}) - 
\epsilon_{di} \frac{\partial B_{di}}{\partial T_d},
\end{eqnarray}
where $k=N\!F+1$ is the index of $T$ in the state vector, and
\begin{eqnarray}
\alpha_{di} &=& (f_{di}J_{di} - f_{d-1}J_{d-1})/(\Delta\tau_{d-1/2,i}\Delta\tau_{di}),\\
\gamma_{di} &=& (f_{di}J_{di} - f_{d+1}J_{d+1})/(\Delta\tau_{d+1/2,i}\Delta\tau_{di}),\\
\beta_{di}&=&\alpha_{di}+\gamma_{di}, \\
a_{di} &=& \big[\alpha_{di} + (\beta_{di}/2)(\Delta\tau_{d-1/2,i}\Delta\tau_{di}\big]/
\omega_{d-1/2,i}, \\
c_{di} &=& \big[\gamma_{di} + (\beta_{di}/2)(\Delta\tau_{d+1/2,i}\Delta\tau_{di}\big]/
\omega_{d+1/2,i},\ \ \ \ 
\end{eqnarray}
where $\omega_{d\pm 1/2} \equiv \omega_d + \omega_{d\pm 1}$.
The right-hand side vector is given by
\begin{equation}
L_{di} = -\beta_{di} -\epsilon_{di}(J_{di} - B_{di}),
\end{equation}
Linearization of the boundary conditions and the radiative/convective
equilibrium equation is analogous

\section{Evaluation of the thermodynamic quantities}

The adiabatic gradient and other thermodynamic quantities can be evaluated using
either the internal energy ($E$), or the entropy ($S$).

When using the internal energy, the corresponding expressions are
\begin{equation}
\nabla_{\rm ad} = \left(\frac{\partial\ln T}{\partial\ln P}\right)_{\!\!S} =
- \frac{P}{\rho c_P T} \left(\frac{\partial\ln \rho}{\partial\ln T}\right)_{\!\!P},
\end{equation}
where the specific heat is given by
\begin{equation}
c_P = \left(\frac{\partial E}{\partial T}\right)_{\!P} - \frac{P}{\rho^2}
\left(\frac{\partial \rho}{\partial T}\right)_{\!P} ,
\end{equation}
and
\begin{equation}
\left(\frac{\partial\ln \rho}{\partial\ln T}\right)_{\!\!P} =\frac{T}{\rho}
\left(\frac{\partial \rho}{\partial T}\right)_{\!P} .
\end{equation}
The internal energy is evaluated as
\begin{equation}
\frac{E}{kT} = \frac{3}{2} + \sum_j N_j\left(\frac{d \ln U_j}{d\ln T}\right),
\end{equation}
where $N_j$ and $U_j$ are the number density and  the partition function of 
species $j$, respectively. The summation is carried over all species.

When using entropy, one has
\begin{equation}
\nabla_{\rm ad} = - \left(\frac{\partial S}{\partial T}\right)_{\!\!P}\Bigg/
\left(\frac{\partial S}{\partial P}\right)_{\!\!T} \frac{P}{T},
\end{equation}
and
\begin{equation}
c_P = -\frac{P}{\rho T}\left(\frac{\partial\ln \rho}{\partial\ln T}\right)_{\!\!P}\Bigg/
\nabla_{\rm ad}
\end{equation}
The entropy is given by
\begin{equation}
S/k = \sum_j N_j [1+\ln(U_j/N_j)] +E/kT.
\end{equation}
All derivatives are evaluated numerically.


\section{Construction of the initial gray model}
\label{gray}
The procedure to construct the initial gray model is very similar to
that described by Kurucz (1970).

First, one sets up a grid of Rosseland optical depths, usually as logarithmically
equidistant between $\tau_1$ and $\tau_{N\!D}$, which are input parameters of
the model. These are typically chosen  as $\tau_1 \approx 10^{-7}$ and 
$\tau_{N\!D} \approx 10^2$. The temperature is a known function of the Rosseland optical
depth, see \S\,\ref{graymu},
\begin{equation}
\label{tgray}
T^4(\tau) = (3/4) T_{\rm eff}^4 [\tau + q(\tau)].+ (\pi/\sigma_{\!R})H^{\rm ext}
\end{equation}
where $q(\tau)$ is the Hopf function, and $H^{\rm ext} = \int_0^\infty H_\nu^{\rm ext} d\nu$
is the frequency-integrated external irradiation flux.

The hydrostatic equilibrium equation is written as
\begin{equation}
\label{hetau}
\frac{d\ln P}{d\ln\tau} = \frac{g\tau}{\chi_{\!R} P},
\end{equation}
because $\tau$ and $P$ span many orders of magnitude, so it is advantageous to 
integrate the equation for logarithms. $\chi_{\!R}$ is the Rosseland mean opacity.

One then proceeds to solving Eq.  (\ref{hetau}) from the top of the atmosphere
to the bottom. At the first depth point, $\tau_1$, one makes a first estimate of
the Rosseland mean opacity, $\chi_{\!R,1}$, and assumes it is 
constant from this
point upward. Using the boundary condition $P(0)=0$, one obtains
the first estimate of the pressure $P_1$ as
\begin{equation}
\label{pr1}
P_1 = (g/\chi_{\!R,1}) \tau_1.
\end{equation}

Having an estimate of the pressure,
one uses the following procedure which is valid for every depth point $d$:
From known temperature $T(\tau_d)$, given by Eq. (\ref{tgray}),  one 
computes monochromatic opacities, and, by integrating over frequency, 
the new value of the Rosseland mean opacity $\chi_{\!R}$.
We will refer to this procedure as $P\!\rightarrow\chi_{\!R}$.
With the new value of $\chi_{\!R}$, one returns to Eq. (\ref{pr1}), evaluates an
improved estimate of $P_1$, and repeats the procedure 
$P\!\rightarrow\chi_{\!R}$
until convergence. Once this is done, one proceeds to the subsequent depth point.

For the next three depth points, $d=2,\ldots,4$,
one obtains the first estimate (a predictor step)
of the total pressure is:
\begin{equation}
\ln P_d^{\rm pred} = \ln P_{d-1} + \Delta\!\ln P_{d-1} ,
\end{equation}
which is followed by a $P\!\rightarrow\chi_{\!R}$ procedure, and with the new 
$\chi_{\!R}$ one goes to the corrector step,
\begin{equation}
\ln P_d = (\ln P_{d}^{\rm pred} + 2 \ln P_{d-1} + \Delta\!\ln P_{d} +  \Delta\!\ln P_{d-1})/3,
\end{equation}
where
\begin{equation}
\Delta\!\ln P_d = \frac{g\tau_d}{\chi_{R,d} P_d} (\ln\tau_d- \ln\tau_{d-1}).
\end{equation}
For the subsequent depth points, one uses the Hamming's predictor-corrector
scheme (see Kurucz 1970; Eqs. 4.17 and 4.18), where the predictor step is
\begin{equation}
\ln P_d = (3\ln P_{d-4} + 8 \ln P_{d-1}- 4\Delta\!\ln P_{d-2} + 8 \Delta\!\ln P_{d-3})/3,
\end{equation}
and the corrector step
\begin{eqnarray}
\ln P_d = (126\ln P_{d-1} -14 \ln P_{d-3}+ 9 \ln P_{d-4} 
+42 \Delta\!\ln P_{d} \nonumber \\
+ 108 \Delta\!\ln P_{d-1}-54\Delta\!\ln P_{d-2} + 24 \Delta\!\ln P_{d-3})/121.
\end{eqnarray}
After completing the above procedure for all depths, one constructs the
column mass scale, which will subsequently be used as the basic depth
scale, as
\begin{equation}
m_d = P_d /g.
\end{equation}
When convection is taken into account, one first computes the radiative gradient
of temperature,
\begin{equation}
\nabla_d = \frac{(T_d - T_{d-1})} {(P_d - P_{d-1})}\frac {(P_d + P_{d-1})}{(T_d + T_{d-1})},
\end{equation}
and compares to the adiabatic gradient, $\nabla_{\rm add}$. 
If $\nabla_{\rm rad} > \nabla_{\rm add}$, the criterion for stability against 
convection is violated, one determines the true 
gradient $\nabla$, where $\nabla_\mathrm{ad}\leq\nabla\leq\nabla_\mathrm{rad}$, 
that gives the correct total, radiative plus convective, flux.
If the instability occurs deep enough for the
diffusion approximation to be valid, then $(F_\mathrm{rad}/F)=
(\nabla/\nabla_\mathrm{ad})$, and the energy balance equation reads (see Hubeny \&
Mihalas 2014, \S\,17.4),
\begin{equation}
\label{congr}
\mathcal{A}\big(\nabla-\nabla_\mathrm{el}\big)^{3/2}=\nabla_\mathrm{rad}-\nabla,
\end{equation}
where 
\begin{equation}
\mathcal{A} = (\nabla_\mathrm{rad}/\sigma_\mathrm{R}T_\mathrm{eff}^4) 
(gQH_P/32)^{1/2}(\rho c_P T) (\ell/H_P)^2 .
\end{equation}
We see that $\mathcal{A}$ depends only on local variables. Adding
$\big(\nabla-\nabla_\mathrm{el}\big)+\big(\nabla_\mathrm{el}
-\nabla_\mathrm{ad}\big)$ to both sides of (\ref{congr}), and using
the expression $\nabla_{\rm el} - \nabla_{\rm ad}= B \sqrt{\nabla-\nabla_{\rm el}}$,
where $B$ is given by Eq. (\ref{convb}), to eliminate
$\big(\nabla_\mathrm{el}-\nabla_\mathrm{ad}\big)$, we obtain a cubic equation
for $x\equiv\big(\nabla-\nabla_\mathrm{el}\big)^{1/2}$, namely
\begin{equation}
\mathcal{A}\big(\nabla-\nabla_\mathrm{el}\big)^{3/2}+\big(\nabla-\nabla_\mathrm{el}\big)+
B\big(\nabla-\nabla_\mathrm{el})^{1/2}=\big(\nabla_\mathrm{rad}-\nabla_\mathrm{ad}\big).
\end{equation}
or
\begin{equation}
\mathcal{A}x^3+x^2+Bx= \big(\nabla_\mathrm{rad}-\nabla_\mathrm{ad}\big),
\end{equation}
which can be solved numerically for the root $x_0$. We thus obtain the true
gradient $\nabla=\nabla_\mathrm{ad}+\mathcal{B}x_0+x_0^2$, and can proceed with
the integration, now regarding $T$ as a function of $P$ and the logarithmic gradient
$\nabla$.


\bsp	
\label{lastpage}
\end{document}